

\input harvmac.tex

\def\vt#1#2#3{ {\vartheta[{#1 \atop  #2}](#3\vert \tau)} }

\def\makeblankbox#1#2{\hbox{\lower\dp0\vbox{\hidehrule{#1}{#2}%
   \kern -#1
   \hbox to \wd0{\hidevrule{#1}{#2}%
      \raise\ht0\vbox to #1{}
      \lower\dp0\vtop to #1{}
      \hfil\hidevrule{#2}{#1}}%
   \kern-#1\hidehrule{#2}{#1}}}%
}%
\def\hidehrule#1#2{\kern-#1\hrule height#1 depth#2 \kern-#2}%
\def\hidevrule#1#2{\kern-#1{\dimen0=#1\advance\dimen0 by #2\vrule
    width\dimen0}\kern-#2}%
\def\openbox{\ht0=1.2mm \dp0=1.2mm \wd0=2.4mm  \raise 2.75pt
\makeblankbox {.25pt} {.25pt}  }

\def\bun#1/#2{\leavevmode
   \kern.1em \raise .5ex \hbox{\the\scriptfont0 #1}%
   \kern-.1em $/$%
   \kern-.15em \lower .25ex \hbox{\the\scriptfont0 #2}%
}

\def\opensquare{\ht0=3.4mm \dp0=3.4mm \wd0=6.8mm  \raise 2.7pt
\makeblankbox {.25pt} {.25pt}  }


\def\sector#1#2{\ {\scriptstyle #1}\hskip 1mm
\mathop{\opensquare}\limits_{\lower 1mm\hbox{$\scriptstyle#2$}}\hskip 1mm}

\def\tsector#1#2{\ {\scriptstyle #1}\hskip 1mm
\mathop{\opensquare}\limits_{\lower 1mm\hbox{$\scriptstyle#2$}}^\sim\hskip 1mm}


\def\CN{{\cal N}}
\def\CZ{{\cal Z}}
\def\g{\gamma}

\def\Ch{{{\rm Ch}}}

\def\IZ{{\bf Z}}
\def\IC{{\bf C}}
\def\IR{{\bf R}}
\def\l{\ell}
\def\mod{{\rm mod}}

\def\p{\partial}
\def\pb{{\bar\partial}}
\def\CS{{\cal S}}

\def\STr{{\rm STr}}
\def\SCh{{\rm SCh}}
\def\SChm{\SCh_{{\rm m}}}
\def\rhom{\rho_{{\rm m}}}
\def\tl{\tilde }
\def\vt#1#2#3{ {\vartheta[{#1 \atop  #2}](#3\vert \tau)} }

\def\cag{{\CA_{\gamma}} }

\def\IH{{\bf H}}
\def\SCh{{\rm SCh} }
\def\qt{{\tilde q}}
\def\syms{ {\rm Sym}^N(\CS)  }

\def\zb{\bar{z}}

\def\A{{A}}



\overfullrule=0pt

\font\manual=manfnt \def\dbend{\lower3.5pt\hbox{\manual\char127}}

\def\cf{{\it c.f.}}

\def\sst{\scriptscriptstyle}

\def\frac#1#2{{#1\over#2}}

\def\half{\frac12}
\def\hf{{\textstyle\half}}
\def\d{\partial}
\def\p{\partial}

\def\inbar{\,\vrule height1.5ex width.4pt depth0pt}
\def\IR{\relax{\rm I\kern-.18em R}}
\def\IC{\relax\hbox{$\inbar\kern-.3em{\rm C}$}}
\def\IQ{\relax\hbox{$\inbar\kern-.3em{\rm Q}$}}
\def\IH{\relax{\rm I\kern-.18em H}}
\def\IN{\relax{\rm I\kern-.18em N}}
\def\IP{\relax{\rm I\kern-.18em P}}
\font\cmss=cmss10
\font\cmsss=cmss10 at 7pt
\def\IZ{\relax\ifmmode\mathchoice
{\hbox{\cmss Z\kern-.4em Z}}{\hbox{\cmss Z\kern-.4em Z}}
{\lower.9pt\hbox{\cmsss Z\kern-.4em Z}}
{\lower1.2pt\hbox{\cmsss Z\kern-.4em Z}}\else{\cmss Z\kern-.4em
Z}\fi}
\def\Z{{\IZ}}

\catcode`\@=11
\def\slash#1{\mathord{\mathpalette\c@ncel{#1}}}
\def\underrel#1\over#2{\mathrel{\mathop{\kern\z@#1}\limits_{#2}}}

\catcode`\@=12
%

%
%
\def\ket#1{|#1\rangle}

\def\Tr{{\rm Tr}}
\def\mod{{\rm mod}}

\def\exp{{\rm exp}}

\def\dim{{\mathop{\rm dim}}}

%
%
 \def\CA{{\cal A}}
 
\def\CC{{\cal C}} 

 \def\CG{{\cal G}}
 \def\CH{{\cal H}}
\def\II{{\cal I}}

 \def\CN{{\cal N}}
 \def\CO{{\cal O}}

 \def\CS{{\cal S}}
 \def\CT{{\cal T}}

 \def\CW{{\cal W}}

 \def\CZ{{\cal Z}}

%
%
\def\unlockat{\catcode`\@=11}
\def\lockat{\catcode`\@=12}
\unlockat
\def\newsec#1{\global\advance\secno by1\message{(\the\secno. #1)}
\global\subsecno=0\global\subsubsecno=0\eqnres@t\noindent
{\bf\the\secno. #1}
\writetoca{{\secsym} {#1}}\par\nobreak\medskip\nobreak}
\global\newcount\subsecno \global\subsecno=0
\def\subsec#1{\global\advance\subsecno
by1\message{(\secsym\the\subsecno. #1)}
\ifnum\lastpenalty>9000\else\bigbreak\fi\global\subsubsecno=0
\noindent{\it\secsym\the\subsecno. #1}
\writetoca{\string\quad {\secsym\the\subsecno.} {#1}}
\par\nobreak\medskip\nobreak}
\global\newcount\subsubsecno \global\subsubsecno=0
\def\subsubsec#1{\global\advance\subsubsecno by1
\message{(\secsym\the\subsecno.\the\subsubsecno. #1)}
\ifnum\lastpenalty>9000\else\bigbreak\fi
\noindent\quad{\secsym\the\subsecno.\the\subsubsecno.}{#1}
\writetoca{\string\qquad{\secsym\the\subsecno.\the\subsubsecno.}{#1}}
\par\nobreak\medskip\nobreak}
\def\subsubseclab#1{\DefWarn#1\xdef
#1{\noexpand\hyperref{}{subsubsection}%
{\secsym\the\subsecno.\the\subsubsecno}%
{\secsym\the\subsecno.\the\subsubsecno}}%
\writedef{#1\leftbracket#1}\wrlabeL{#1=#1}}
\lockat
%

\def\cag{{\CA_\gamma}}



\def\unlockat{\catcode`\@=11}
\def\lockat{\catcode`\@=12}

\unlockat

\def\newsec#1{\global\advance\secno by1\message{(\the\secno. #1)}
\global\subsecno=0\global\subsubsecno=0\eqnres@t\noindent
{\bf\the\secno. #1}
\writetoca{{\secsym} {#1}}\par\nobreak\medskip\nobreak}
\global\newcount\subsecno \global\subsecno=0
\def\subsec#1{\global\advance\subsecno
by1\message{(\secsym\the\subsecno. #1)}
\ifnum\lastpenalty>9000\else\bigbreak\fi\global\subsubsecno=0
\noindent{\it\secsym\the\subsecno. #1}
\writetoca{\string\quad {\secsym\the\subsecno.} {#1}}
\par\nobreak\medskip\nobreak}
\global\newcount\subsubsecno \global\subsubsecno=0
\def\subsubsec#1{\global\advance\subsubsecno by1
\message{(\secsym\the\subsecno.\the\subsubsecno. #1)}
\ifnum\lastpenalty>9000\else\bigbreak\fi
\noindent\quad{\secsym\the\subsecno.\the\subsubsecno.}{#1}
\writetoca{\string\qquad{\secsym\the\subsecno.\the\subsubsecno.}{#1}}
\par\nobreak\medskip\nobreak}

\def\subsubseclab#1{\DefWarn#1\xdef
#1{\noexpand\hyperref{}{subsubsection}%
{\secsym\the\subsecno.\the\subsubsecno}%
{\secsym\the\subsecno.\the\subsubsecno}}%
\writedef{#1\leftbracket#1}\wrlabeL{#1=#1}}
\lockat



\font\cmss=cmss10 \font\cmsss=cmss10 at 7pt

\def\IB{\relax\hbox{$\inbar\kern-.3em{\rm B}$}}
\def\IC{\relax\hbox{$\inbar\kern-.3em{\rm C}$}}
\def\IQ{\relax\hbox{$\inbar\kern-.3em{\rm Q}$}}
\def\ID{\relax\hbox{$\inbar\kern-.3em{\rm D}$}}
\def\IE{\relax\hbox{$\inbar\kern-.3em{\rm E}$}}
\def\IF{\relax\hbox{$\inbar\kern-.3em{\rm F}$}}
\def\IG{\relax\hbox{$\inbar\kern-.3em{\rm G}$}}
\def\IGa{\relax\hbox{${\rm I}\kern-.18em\Gamma$}}
\def\IH{\relax{\rm I\kern-.18em H}}
\def\IK{\relax{\rm I\kern-.18em K}}
\def\IL{\relax{\rm I\kern-.18em L}}
\def\IP{\relax{\rm I\kern-.18em P}}
\def\IR{\relax{\rm I\kern-.18em R}}
\def\Z{\relax\ifmmode\mathchoice
{\hbox{\cmss Z\kern-.4em Z}}{\hbox{\cmss Z\kern-.4em Z}}
{\lower.9pt\hbox{\cmsss Z\kern-.4em Z}}
{\lower1.2pt\hbox{\cmsss Z\kern-.4em Z}}\else{\cmss Z\kern-.4em
Z}\fi}
\def\II{\relax{\rm I\kern-.18em I}}

\def\S{{\bf S}}

\def\hf{{1\over 2}}

\def\CA {{\cal A}}

\def\CC {{\cal C}}

\def\CG {{\cal G}}
\def\CH {{\cal H}}

\def\CN {{\cal N}}
\def\CO {{\cal O}}

\def\CS {{\cal S}}
\def\CT {{\cal T}}

\def\CW {{\cal W}}

\def\CZ {{\cal Z}}


\def\p{\partial}
\def\pb{\bar{\partial}}


\def\zb {\bar{z}}


\def\Tr{{\rm Tr}}

\def\p{\partial}
\def\pb{\bar{\partial}}

\def\inbar{\,\vrule height1.5ex width.4pt depth0pt}

\def\l{{\ell}}

\def\a{\alpha}
\def\b{\beta}
\def\g{\gamma}
\def\d{\delta}

\def\bar{\overline}

\def\example#1{\bgroup\narrower\footnotefont\baselineskip\footskip\bigbreak
\hrule\medskip\nobreak\noindent {\bf Example}. {\it #1\/}\par\nobreak}
\def\endexample{\medskip\nobreak\hrule\bigbreak\egroup}


\lref\AlvarezGaumeVM{
L.~Alvarez-Gaume, J.~B.~Bost, G.~W.~Moore, P.~Nelson and C.~Vafa,
``Bosonization On Higher Genus Riemann Surfaces,''
Commun.\ Math.\ Phys.\  {\bf 112}, 503 (1987).
}

\lref\bowditch{B.H. Bowditch, Geometrical finiteness for hyperbolic groups,
J. Funct. Anal. {\bf 113} (1993)245}

\lref\deBoerRH{
J.~de Boer, A.~Pasquinucci and K.~Skenderis,
``AdS/CFT dualities involving large 2d N = 4 superconformal symmetry,''
Adv.\ Theor.\ Math.\ Phys.\  {\bf 3}, 577 (1999)
[arXiv:hep-th/9904073].
}

\lref\cvl{M. Cveti\'c and F. Larsen, ``Near Horizon Geometry
of Rotating Black Holes in Five Dimensions,''
hep-th/9805097}
\lref\breckenridge{J.C. Breckenridge,  D.A. Lowe,  R.C. Myers,  A.W. Peet,
A. Strominger, C. Vafa,
``Macroscopic and Microscopic Entropy of Near-Extremal Spinning Black Holes,''
hep-th/9603078; Phys.Lett. B381 (1996) 423-426}
\lref\cveticyoum{M. Cvetic and D. Youm,
``General Rotating Five Dimensional Black Holes of Toroidally Compactified
Heterotic String,''
hep-th/9603100; Nucl.Phys. B476 (1996) 118-132}

\lref\DijkgraafXW{
R.~Dijkgraaf, G.~W.~Moore, E.~Verlinde and H.~Verlinde,
``Elliptic genera of symmetric products and second quantized strings,''
Commun.\ Math.\ Phys.\  {\bf 185}, 197 (1997)
[arXiv:hep-th/9608096].
}
\lref\DijkgraafZA{
R.~Dijkgraaf,
``Discrete torsion and symmetric products,''
arXiv:hep-th/9912101.
}

\lref\GauntlettKC{
J.~P.~Gauntlett, R.~C.~Myers and P.~K.~Townsend,
``Supersymmetry of rotating branes,''
Phys.\ Rev.\ D {\bf 59}, 025001 (1999)
[arXiv:hep-th/9809065].
}

\lref\PTone{J.~L.~Petersen and A.~Taormina,
``Characters Of The N=4 Superconformal Algebra With Two Central Extensions,''
Nucl.\ Phys.\ B {\bf 331}, 556 (1990).}

\lref\PTtwo{J.~L.~Petersen and A.~Taormina,
``Characters Of The N=4 Superconformal Algebra With Two Central
Extensions: 2. Massless Representations,''
Nucl.\ Phys.\ B {\bf 333}, 833 (1990).}

\lref\deBoer{J.~de Boer, A.~Pasquinucci and K.~Skenderis,
``AdS/CFT dualities involving large 2d N = 4 superconformal symmetry,''
Adv.\ Theor.\ Math.\ Phys.\  {\bf 3}, 577 (1999); hep-th/9904073.}

\lref\ManinHN{
Y.~I.~Manin and M.~Marcolli,
``Holography principle and arithmetic of algebraic curves,''
Adv.\ Theor.\ Math.\ Phys.\  {\bf 5}, 617 (2002)
[arXiv:hep-th/0201036].
}

\lref\PT{J.~L.~Petersen and A.~Taormina,
``Characters Of The N=4 Superconformal Algebra With Two Central Extensions,''
Nucl.\ Phys.\ B {\bf 331}, 556 (1990);
``Characters Of The N=4 Superconformal Algebra With Two Central
Extensions: 2. Massless Representations,''
Nucl.\ Phys.\ B {\bf 333}, 833 (1990).}

\lref\GPTVP{M.~Gunaydin, J.~L.~Petersen, A.~Taormina and A.~Van Proeyen,
``On The Unitary Representations Of A Class Of N=4 Superconformal Algebras,''
Nucl.\ Phys.\ B {\bf 322}, 402 (1989).}

\lref\OoguriDS{
H.~Ooguri, J.~L.~Petersen and A.~Taormina,
``Modular invariant partition functions for the 
doubly extended N=4 superconformal algebras,''
Nucl.\ Phys.\ B {\bf 368}, 611 (1992).
}
\lref\MaldacenaBP{
J.~M.~Maldacena, G.~W.~Moore and A.~Strominger,
``Counting BPS black holes in toroidal type II string theory,''
arXiv:hep-th/9903163.
}
\lref\MaldacenaBW{
J.~M.~Maldacena and A.~Strominger,
``AdS(3) black holes and a stringy exclusion principle,''
JHEP {\bf 9812}, 005 (1998)
[arXiv:hep-th/9804085].
}

\lref\DijkgraafFQ{
R.~Dijkgraaf, J.~M.~Maldacena, G.~W.~Moore and E.~Verlinde,
``A black hole farey tail,''
arXiv:hep-th/0005003.
}

\lref\SevrinEW{
A.~Sevrin, W.~Troost and A.~Van Proeyen, ``Superconformal Algebras
In Two-Dimensions With N=4,'' Phys.\ Lett.\ B {\bf 208}, 447 (1988). }

\lref\StromingerEQ{
A.~Strominger,
``Black hole entropy from near-horizon microstates,''
JHEP {\bf 9802}, 009 (1998)
[arXiv:hep-th/9712251].
}

\lref\deBoerIP{
J.~de Boer,
``Six-dimensional supergravity on $S^3 \times AdS_3$ 
and 2d conformal field  theory,''
Nucl.\ Phys.\ B {\bf 548}, 139 (1999)
[arXiv:hep-th/9806104].
}

\lref\deBoerUS{
J.~de Boer,
``Large N Elliptic Genus and AdS/CFT Correspondence,''
JHEP {\bf 9905}, 017 (1999)
[arXiv:hep-th/9812240].
}

\lref\ElitzurMM{S.~Elitzur, O.~Feinerman, A.~Giveon and D.~Tsabar,
``String theory on AdS(3) x S(3) x S(3) x S(1),''
Phys.\ Lett.\ B {\bf 449}, 180 (1999); hep-th/9811245.}

\lref\CecottiQH{
S.~Cecotti, P.~Fendley, K.~A.~Intriligator and C.~Vafa,
``A New supersymmetric index,''
Nucl.\ Phys.\ B {\bf 386}, 405 (1992)
[arXiv:hep-th/9204102].
}

\lref\GunaydinZW{
M.~Gunaydin,
``On The Chiral Rings In N=2 And N=4 Superconformal Algebras,''
Int.\ J.\ Mod.\ Phys.\ A {\bf 8}, 301 (1993)
[arXiv:hep-th/9207023].
}

\lref\ez{M. Eichler and D. Zagier, {\it The theory of
Jacobi forms}, Birkh\"auser 1985}

%
\lref\GukovYM{
S.~Gukov, E.~Martinec, G.~Moore and A.~Strominger,
``The search for a holographic dual to 
$AdS_3 \times S^3 \times S^3 \times S^1$,''
arXiv:hep-th/0403090.
}

%
\lref\GukovID{
S.~Gukov, E.~Martinec, G.~Moore and A.~Strominger,
``Chern-Simons gauge theory and the $AdS_3/CFT_2$ correspondence,''
arXiv:hep-th/0403225.
}

\lref\OpfermannQD{
A.~Opfermann and G.~Papadopoulos,
arXiv:math-ph/9807026.
}

\lref\Srep{K.~Schoutens, ``A Nonlinear Representation Of The D = 2
SO(4) Extended Superconformal Algebra,'' Phys.\ Lett.\ B {\bf
194}, 75 (1987).}

\lref\KawaiJK{
T.~Kawai, Y.~Yamada and S.~K.~Yang,
``Elliptic genera and N=2 superconformal field theory,''
Nucl.\ Phys.\ B {\bf 414}, 191 (1994)
[arXiv:hep-th/9306096].
}

\lref\GPTVP{M.~Gunaydin, J.~L.~Petersen, A.~Taormina and A.~Van Proeyen,
``On The Unitary Representations Of A Class Of N=4 Superconformal Algebras,''
Nucl.\ Phys.\ B {\bf 322}, 402 (1989).}

\lref\witteneg{E. Witten, ``Elliptic Genera and
Quantum Field Theory,'' Commun. Math. Phys. {\bf 109}(1987)525;
``The index of the Dirac operator in loop space,'' Proceedings of
the conference on elliptic curves and modular forms in algebraic
topology, Princeton NJ, 1986.}

\lref\AlvarezWG{
O.~Alvarez, T.~P.~Killingback, M.~L.~Mangano and P.~Windey,
``String Theory And Loop Space Index Theorems,''
Commun.\ Math.\ Phys.\  {\bf 111}, 1 (1987).
}
\lref\AlvarezDE{
O.~Alvarez, T.~P.~Killingback, M.~L.~Mangano and P.~Windey,
``The Dirac-Ramond Operator In String Theory And Loop Space Index Theorems,''
UCB-PTH-87/11
{\it Invited talk presented at the Irvine Conf. 
on Non-Perturbative Methods in Physics, Irvine, Calif., Jan 5-9, 1987}
}
\lref\WindeyAR{
P.~Windey,
``The New Loop Space Index Theorems And String Theory,''
 {\it Lectures given at 25th Ettore Majorana Summer School 
for Subnuclear Physics, Erice, Italy, Aug 6-14, 1987}
}

\lref\MooreFG{
G.~W.~Moore,
``Les Houches lectures on strings and arithmetic,''
arXiv:hep-th/0401049.
}
\lref\DijkgraafFQ{
R.~Dijkgraaf, J.~M.~Maldacena, G.~W.~Moore and E.~Verlinde,
``A black hole farey tail,''
arXiv:hep-th/0005003.
}
\lref\StromingerSH{
A.~Strominger and C.~Vafa,
``Microscopic Origin of the Bekenstein-Hawking Entropy,''
Phys.\ Lett.\ B {\bf 379}, 99 (1996)
[arXiv:hep-th/9601029].
}

\lref\Ivanov{E.A. Ivanov, S.O. Krivonos, ``N=4 super-Liouville equation'',
J.Phys. A: Math. Gen., 17 (1984) L671; ``N=4 superextension of the
Liouville equation with quaternionic structure'', Teor. Mat. Fiz.
63 (1985) 230 [Theor. Math. Phys. 63 (1985) 477];
E.A. Ivanov, S.O. Krivonos, V.M. Leviant, ``A new class of
superconformal sigma models with the Wess-Zumino action'',
Nucl. Phys. B304 (1988) 601;
E.A. Ivanov, S.O. Krivonos, V.M. Leviant, ``Quantum N=3, N=4
superconformal WZW sigma models'', Phys. Lett. B215 (1988) 689; B221
(1989) 432E.}

\lref\LuninJY{
O.~Lunin and S.~D.~Mathur,
Nucl.\ Phys.\ B {\bf 623}, 342 (2002)
[arXiv:hep-th/0109154].
}

\lref\LercheCA{
W.~Lerche and N.~P.~Warner,
``Index Theorems In N=2 Superconformal Theories,''
Phys.\ Lett.\ B {\bf 205}, 471 (1988).
}

\lref\LercheQK{
W.~Lerche, B.~E.~W.~Nilsson, A.~N.~Schellekens and N.~P.~Warner,
``Anomaly Cancelling Terms From The Elliptic Genus,''
Nucl.\ Phys.\ B {\bf 299}, 91 (1988).
}

\lref\PilchGS{
K.~Pilch and N.~P.~Warner,
``String Structures And The Index Of The Dirac-Ramond Operator On Orbifolds,''
Commun.\ Math.\ Phys.\  {\bf 115}, 191 (1988).
}

\lref\PilchEN{
K.~Pilch, A.~N.~Schellekens and N.~P.~Warner,
``Path Integral Calculation Of String Anomalies,''
Nucl.\ Phys.\ B {\bf 287}, 362 (1987).
}

\lref\SchellekensYJ{
A.~N.~Schellekens and N.~P.~Warner,
``Anomaly Cancellation And Selfdual Lattices,''
Phys.\ Lett.\ B {\bf 181}, 339 (1986).
}

\lref\SchellekensYI{
A.~N.~Schellekens and N.~P.~Warner,
``Anomalies And Modular Invariance In String Theory,''
Phys.\ Lett.\ B {\bf 177}, 317 (1986).
}

%
\Title{\vbox{\baselineskip12pt
\hbox{hep-th/0404023}
\hbox{HUTP-04/A015}
\hbox{EFI-04-10}
}}
{\vbox{\centerline {An Index for $2D$ field theories with large $\CN=4$ }
\centerline{superconformal symmetry } }}
\centerline{Sergei Gukov\footnote{$^*$}{\it Jefferson Physical
Laboratory, Harvard University, Cambridge, MA 02138},
Emil Martinec\footnote{$^{**}$}{\it University of Chicago,
Chicago, IL 60637},
Gregory Moore\footnote{$^\dagger$}{\it Department of Physics and
Astronomy, Rutgers University, Piscataway, NJ 08855} and Andrew
Strominger$^*$} \vskip.1in \vskip.1in \centerline{\bf Abstract}
\noindent
We consider families of theories with large $\CN=4$ superconformal
 symmetry. We define an index
generalizing the elliptic genus of theories with $\CN=2$ symmetry.
In contrast to the $\CN=2$ case, the new index constrains part of the
non-BPS spectrum. Motivated by aspects of the AdS/CFT correspondence
we study the index in the examples of symmetric product theories.
We give a physical interpretation of the Hecke operators which
appear in the expressions for partition functions of such theories.
Finally, we  compute the index for a nontrivial example of a symmetric product
theory.

\Date{April 2, 2004}

\listtoc\writetoc


\newsec{Introduction and Summary}

The elliptic genus of an $\CN=2$ superconformal field
theory is a powerful tool in analyzing the field content
of subtle conformal field theories (such as CY sigma models)
that depend on parameters
\refs{\SchellekensYI,\SchellekensYJ,\PilchEN,\witteneg,
\AlvarezWG,\AlvarezDE,\WindeyAR,\PilchGS,\LercheQK,\LercheCA,\KawaiJK}.
The reason it is useful is that
 it is invariant
under deformations of parameters that preserve the $\CN=2$ supersymmetry.
Usually, one can deform parameters to a region where the genus
can be evaluated explicitly. One then obtains some nontrivial information
about the CFT for all values of parameters. In this paper we will
describe an analog of the elliptic genus for theories with
a larger superconformal symmetry, namely, the maximal, or
large $\CN=4$ superconformal symmetry $\cag$ discovered in \SevrinEW.
Some background on the large superconformal algebra
can be found in \refs{\GPTVP,\PTone,\PTtwo,\OoguriDS, \deBoerRH,\GukovYM}.
We use the conventions described in \GukovYM.

In conformal field theories with $\cag$ symmetry one can expect
that the larger amount of symmetry leads to more control of the
spectrum. Here we show that this is indeed the case. In equations
$(3.3)$ and $(3.4)$  below we define an index for theories with
$\cag$ symmetry which is  invariant under deformations preserving
$\cag$ symmetry. The index satisfies some novel properties
reflecting the greater control on the spectrum. For example, it is
not holomorphic, and does not just count BPS states.

A natural operation on conformal field theories is the
symmetric product orbifold. This operation has proven to be
of great importance in Matrix string theory and in the AdS/CFT
correspondence. It is therefore natural to study the $\cag$ index
for such theories. If $\CC_0$ is a conformal field theory with
$\cag$ symmetry then it turns out that there is a
remarkably simple formula for the index of
 ${\rm Sym}^N(\CC_0)$.
This is equation $(4.6)$ below, which expresses the index in terms
of a Hecke transform of the partition function of $\CC_0$. The
occurance of Hecke operators in formulae for partition functions
of symmetric product theories has been noted previously
\DijkgraafXW. In section 4.2 below   we give a simple physical
interpretation to some aspects of the Hecke algebra in terms of
``short string'' and ``long string'' operators.

In section 5  we  describe the computation of the index in a
nontrivial example, ${\rm Sym}^N(\CS)$, where $\CS$ is the
simplest theory with $\cag$ symmetry. We also describe some
partial results for some other theories with $\cag$ symmetry in
section 6. Several definitions and conventions on theta functions
are relegated to the appendices.

This work was motivated by our search for a holographic dual to
type II string theory on $AdS_3 \times \S^3 \times \S^3 \times
\S^1$. Applications to that problem appear in \GukovYM. However,
since the results on the index might be of use in other
applications of superconformal field theory we have written a
separate paper. For other applications of elliptic genera to the
AdS/CFT correspondence see
\refs{\StromingerSH,\deBoerIP,\MaldacenaBP,\DijkgraafFQ,\MooreFG}.
\newsec{Prelimimaries}
In this section we briefly review some aspects of the $\CA_{\g}$
algebra and establish our notation.

 \subsec{The superconformal algebra
$\CA_{\g}$}

Apart from the usual Virasoro algebra, the large $\CN=4$
superconformal algebra $\CA_{\g}$ contains two copies of the
affine $\widehat{SU(2)}$ Lie algebras, at the levels $k^+$ and
$k^-$, respectively. 
We also define $k\equiv k^++k^-$.
The relation between $k^{\pm}$ and the
parameter $\g$ is
\eqn\gviakk{ \g = {k^- \over k^+ + k^-} \ .}
Unitarity implies that the Virasoro central charge is:
\eqn\cviakk{ c= {6 k^+ k^- \over k^+ + k^-}\ . }

The superconformal algebra $\CA_{\g}$ is generated by six affine
$\widehat{SU(2)}$ generators $A^{\pm,i} (z)$, four dimension $3/2$
supersymmetry generators $G^a (z)$, four dimension $1/2$ fields
$Q^a (z)$, a dimension 1 field $U(z)$, and the Virasoro current
$T(z)$. The OPEs with the Virasoro generators, $T_m$, have the
usual form. The remaining OPEs are \refs{\SevrinEW,\Srep}:
\eqn\agope{\eqalign{ & G^a (z) G^b (w) = {2c \over 3} {\d^{ab}
\over (z-w)^3} - {8 \g \a_{ab}^{+,i} A^{+,i} (w) + 8 (1-\g)
\a_{ab}^{-,i} A^{-,i}(w) \over (z-w)^2 } - \cr &~~~~~~~~~~~~~~~ -
{4\g \a_{ab}^{+,i} \p A^{+,i} (w) + 4 (1-\g)\a_{ab}^{-,i} \p
A^{-,i} (w) \over z-w } + {2\d^{ab} L(w) \over z-w} + \ldots, \cr
& A^{\pm, i} (z) A^{\pm ,j} (w) = -{k^{\pm} \d^{ij} \over 2
(z-w)^2} + {\epsilon^{ijk} A^{\pm,k} (w) \over z-w} + \ldots, \cr
& Q^a (z) Q^b (w) = - {(k^+ + k^-) \d^{ab} \over 2 (z-w)} +
\ldots, \cr & U(z) U(w) = - {k^+ + k^- \over 2 (z-w)^2} + \ldots,
\cr & A^{\pm, i} (z) G^a (w) = \mp {2 k^{\pm} \a_{ab}^{\pm,i} Q^b
(w) \over (k^+ + k^-) (z-w)^2 } + {\a_{ab}^{\pm,i} G^b (w) \over
z-w } + \ldots, \cr & A^{\pm,i} (z) Q^a (w) = {\a_{ab}^{\pm,i} Q^b
(w) \over z-w } + \ldots, \cr & Q^a (z) G^b (w) = {2 \a_{ab}^{+,i}
A^{+,i} (w) - 2 \a_{ab}^{-,i} A^{-,i} (w) \over z-w} + {\d^{ab}
U(w) \over z-w} + \ldots, \cr & U(z) G^a(w) = {Q^a (w) \over
(z-w)^2}  + \ldots. }}
$\a_{ab}^{\pm,i}$ here are $4 \times 4$ matrices, which project
onto (anti)self-dual tensors.  Explicitly,
\eqn\thoofteta{ \a^{\pm,i}_{ab} = {1 \over 2} \Big( \pm \d_{ia}
\d_{b0} \mp \d_{ib} \d_{a0} + \epsilon_{iab} \Big) \ .}
They obey $SO(4)$ commutation relations: \eqn\sofourcr{
[\a^{\pm,i} , \a^{\pm,j}] = - \epsilon^{ijk} \a^{\pm k} \quad ,
\quad [\a^{+,i} , \a^{-,j}] = 0 \quad , \quad \{ \a^{\pm,i} ,
\a^{\pm, j} \} = - {1 \over 2} \d^{ij}\ . } It is sometimes useful
to employ spinor notation, where for instance $G^a\to G^{A\dot
A}=\gamma_a^{A\dot A}G^a$ (and $\gamma_a^{A\dot A}$ are Dirac
matrices); $A^{+,i}\to A^{AB}= \tau^{AB}_i A^{+,i}$ (where
$\tau^i$ are Pauli matrices); $A^{-,i}\to A^{\dot A\dot B}=
\tau^{\dot A\dot B}_i A^{-,i}$; and so on. An important subalgebra
of $\CA_\gamma$ is denoted $D(2,1|\alpha)$; here
$\alpha=k^-/k^+=\frac{\gamma}{1-\gamma}$. It is generated (in the
NS sector) by $L_0$, $L_{\pm 1}$, $G^a_{\pm 1/2}$, and $A^{\pm,
i}_0$.


\subsec{Examples of large $\CN=4$ SCFT's}

The simplest example of a large $\CN=4$ theory can be realized as
a theory of a free boson, $\phi$, and four Majorana fermions,
$\psi_a$, $a=0,\ldots,3$.  Specifically, we have 
\refs{\Ivanov,\Srep,\SevrinEW}:
\eqn\cthree{\eqalign{ & T = - {1 \over 2} (\p \phi)^2 - {1 \over
2} \psi^a \p \psi^a \cr & G^a  = - {1 \over 6} i \epsilon^{abcd}
\psi^b  \psi^c  \psi^d - i \psi^a \p \phi \cr & A^{\pm,i} = {i
\over 2} \a^{\pm,i}_{ab} \psi^a \psi^b \cr & Q^a = \psi^a \cr & U
= i \p \phi\ . }}
This theory was called the $\CT_3$ theory in \ElitzurMM, but we
shall herein use the notation $\CS$ for simple.

The CFT $\CS$ belongs to a family of large $\CN=4$ theories,
labeled by a non-negative integer number $\kappa$ \SevrinEW:
\eqn\algeone{ \eqalign{ T &  = - J^0 J^0 - {J^a J^a \over
\kappa+2} - \p \psi^a \psi^a \cr G^a & =2 J^0 \psi^a + {4 \over
\sqrt{\kappa+2}} \a^{+,i}_{ab} J^i \psi^b - {2 \over 3
\sqrt{\kappa+2}} \epsilon_{abcd} \psi^b \psi^c \psi^d\cr A^{-,i} &
= \a_{ab}^{-,i} \psi^a \psi^b \cr A^{+,i} & = \a_{ab}^{+,i} \psi^a
\psi^b + J^i \cr U & = -\sqrt{\kappa+2} J^0 \cr Q^a & =
\sqrt{\kappa+2} \psi^a \cr} }
%
where $J^i$ denote $SU(2)$ currents at level $\kappa$ and
$J^0(z)J^0(w) \sim -\half (z-w)^{-2}$.  We shall denote these
theories $\CS_\kappa$. It is easy to check that \algeone\ indeed
generate the large $\CN=4$ algebra with $k^+ = \kappa+1$ and $k^-
= 1$. In fact, the $U(2)$ level $\kappa$ theory of \SevrinEW\
admits {\it two} distinct large $\CN=4$ algebras. The second
algebra is obtained by the outer automorphism and has $(k^+=1, k^-
= \kappa+1)$:
\eqn\algtwo{ \eqalign{ T &  = - J^0 J^0 - {J^a J^a \over \kappa+2}
- \p \psi^a \psi^a \cr G^a & =2 J^0 \psi^a + {4 \over
\sqrt{\kappa+2}} \a^{+,i}_{ab} J^i \psi^b - {2 \over 3
\sqrt{\kappa+2}} \epsilon_{abcd} \psi^b \psi^c \psi^d\cr A^{-,i} &
= \a_{ab}^{-,i} \psi^a \psi^b +J^i \cr A^{+,i} & = \a_{ab}^{+,i}
\psi^a \psi^b   \cr U & = +\sqrt{\kappa+2} J^0 \cr Q^a & =
-\sqrt{\kappa+2} \psi^a \cr} }
The $c=3$ CFT $\CS=\CS_0$ appears as a special case, $\kappa=0$.

Additional examples of large $\CN=4$ are provided by WZW coset
models $\CW\times U(1)$, where $\CW$ is a gauged WZW model
associated to a quaternionic (Wolf) space. Examples based on
classical groups are $\CW=G/H=\frac{SU(n)}{SU(n-2)\times U(1)}$,
$\frac{SO(n)}{SO(n-4)\times SU(2)}$, and
$\frac{Sp(2n)}{Sp(2n-2)}$. These theories carry large $\CN=4$
supersymmetry, with $k^+=\kappa+1$ and $k^-=\check c_G$; here
$\kappa$ is the level of the bosonic current algebra for the group
$G$ and $\check c_G$ its dual Coxeter number.


\subsec{Unitary representations}

The unitary representations of the superconformal algebra
$\CA_{\g}$ are labeled by the conformal dimension $h$, by the
$SU(2)$ spins $\l^{\pm}$, and by the $U(1)$ charge $u$. Character
expansions can be found in appendix A. The generic {\it long} or
{\it massive} representation has no null vectors under the raising
operators of the algebra. On the other hand, the highest weight
states $\ket{\Omega}^{~}_{\!\cag}$ of {\it short} or {\it
massless} representations have the null vector \GPTVP
\eqn\nullvec{
  \Bigl(G^{+\dot+}_{-1/2}-\frac{2u}{k^++k^-}\,
    Q^{+\dot+}_{-1/2}-{2i(\ell^+-\ell^-)\over k^++k^-}\,Q^{+\dot+}_{-1/2}\Bigr)
    \ket{\Omega}^{~}_{\!\cag}=0\ .
} (We have used the property that $ \ket{\Omega}^{~}_{\!\cag}$ is
a highest weight state for the $SU(2)$ current algebras.) Squaring
this null vector leads to a relation among the spins $\ell^\pm$
and the conformal dimension $h$ \refs{\GPTVP,\PT,\PTtwo,\deBoerRH}
\eqn\unitary{
   h_{\rm short} = {1 \over k^+ + k^-}
    \left( k^- \l^+ + k^+ \l^- + (\l^+ - \l^-)^2 + u^2 \right)\ .
} Unitarity demands that all representations, short or long, lie
at or above this bound: $h\ge h_{\rm short}$; and that the spins
lie in the range \hbox{$\ell^\pm=0,\hf,...,\hf(k^\pm-1)$}. When we
consider $U(1)$ singlets, we shall denote representations by their
labels $(h,\ell^+,\ell^-)$; for short representations with $u=0$
it is sufficient to specify them simply by $(\ell^+,\ell^-)$. The
conformal dimension of short representations is protected, as long
as they do not combine into long ones.


\newsec{A new index for theories with $\cag$ symmetry }

In order to define a theory with $\cag$ symmetry we begin with a
representation $\CH_{RR}$   of the RR sector algebra $\cag_{\rm
left} \oplus \cag_{\rm right}$.  We then define the $NS$ sector
modules using  spectral flow.\foot{Then, if one wishes, one can
impose a GSO projection. Of course, the GSO projected theory is in
general not a representation of  $\cag_{\rm left} \oplus \cag_{\rm
right}$.  } The representation content of the theory is therefore
summarized by the  RR sector   supercharacter: 
\eqn\defnewzee{
  Z(\tau, \omega_+,\omega_-; \bar \tau, \tl \omega_+,\tl \omega_-):=
	{\Tr}_{\CH_{RR} } \Bigl[q^{L_0-c/24}\tilde q^{\tilde L_0-c/24}
	z_+^{2A_0^{+,3}} (-z_-)^{2A_0^{-,3}} \tilde z_+^{2\tilde
	A_0^{+,3}} (-\tilde z_-)^{2\tilde A_0^{-,3}} \Bigr]\ .
} 
Here and hereafter
we denote  $z_\pm = e^{2\pi i \omega_\pm }$ for left-movers and
$\tl z_\pm = e^{2\pi i \tl \omega_\pm }$ for right-movers. We can
take the $\omega$'s to be real. Sometimes we simply write $Z$ for
the partition function \defnewzee, or $Z(\CC)$ if we wish to
emphasize dependence on the conformal field theory $\CC$.


Now \defnewzee\ can be expanded in the supercharacters of the irreducible
representations, defined by
\eqn\defchrs{ \SCh(\rho)(\tau,\omega_+,\omega_- ) = {\Tr}_{\rho}
q^{L_0-c/24} z_+^{2A_0^{+,3}} z_-^{2A_0^{-,3}} (-1)^{2A_0^{-,3}} }
we just write $\SCh(\rho)$ when the arguments are understood.
Explicit formulae for these characters 
have been derived by Peterson and Taormina
and are reproduced in appendix B.
Using the formulae of \PTtwo\ one finds that 
short representations have a character
with a first order zero at $z_+ = z_-$, while long representations have a
character with a second order zero at $z_+ = z_-$.

Thanks to the second order vanishing of the characters of long representations
we can define the {\it left-index} of the CFT $\CC$  by
\eqn\soperat{
I_1(\CC):= - z_+
{d\over dz_-}\biggl\vert_{z_- =  z_+} Z\ .
}
Only short, or BPS representations can contribute on the left. On the
right, long representations might contribute. However, due to the constraint
$h-\bar h = 0 ~\mod~ 1$ the right-moving conformal weights which do contribute
are rigid, and hence $I_1$ is a deformation invariant.

Of course, one could also define a right-index. 
Since we will consider left-right
symmetric theories here this is redundant information. Nevertheless, it is
often useful to define the left-right index:
\eqn\indxone{
I_2(\CC)  := z_+ \tl z_+ {d\over dz_-}{d\over d\tilde z_-} Z
}
where one evaluates at $ z_- =  z_+$, $\tl z_- =  \tl z_+$.

Finally, using the explicit formulae of \PTtwo\ 
it is not difficult to show that
the contribution of the massless R sector rep $(\l^+,\l^-,u)$ to the index is
\eqn\contrib{- z_+
{d\over dz_-}\biggl\vert_{z_- =  z_+}\SCh(\l^+,\l^-,u)=
 (-1)^{2\l^-+1} q^{u^2/k} \Theta^-_{\mu,k}(\omega,\tau)
}
where $z=z_+ = \exp[2\pi i \omega]$, $ \Theta^-_{\mu,k}$ is an odd level
$k$-theta function (see appendix D) and
\eqn\drmu{
\mu = \mu(\rho):= 2(\l^+ + \l^-)-1\ .
}

We would like to make a number of remarks on this key result.

\item{1.} When the highest weight state of a long representation
saturates the BPS bound the representation decomposes
as a sum of two short representations.
The formula relating massive and massless characters
when $h$ satisfies the unitarity bound is (\PTtwo\ eq. 5.7)
\eqn\mssmv{
\Ch(\l^+ , \l^-) + \Ch(\l^+-\half  , \l^-+\half )
=\Ch_{\rm massive}(h,\l^+ , \l^-+\half)
}
and it is a nice check that the sum of the indices vanishes.
If the index vanishes on a linear combination of
characters then using \mssmv\ that combination can be written as
an integral linear combination (with signs) of massive characters.

\item{2.} The unitarity bounds on spins in the R sector are
$1/2 \leq  \l^\pm \leq \half  k^\pm$. 
In \contrib\ we have odd level $k$ theta functions.
There are $k-1$ linearly independent such  odd functions. 
For the unitary range,
$\mu$ takes precisely the values $1, \dots, k-1$.

\item{3.}  The only states which can contribute to the index have
\eqn\condi{ 2(A^{+,3} + A^{-,3} ) = \pm \mu(\rho) ~\mod~ 2k }
\eqn\condii{ L_0 -c/24 = {u^2\over k}  +    {1\over k} \biggl[
(A^{+,3} + A^{-,3} )^2  \biggr]\ . }
(This might appear confusing since the BPS bound is expressed in terms of
representation labels in terms of ${1\over k} (\l^+ + \l^- - 1/2)^2$.
The point is that there is not a unique highest weight vector amongst
the R-sector groundstates, and the representation labels $(\l^+,\l^-)$
do not label the quantum numbers of any particular state. See the character
formulae in appendices A and B.)

\item{4.} The fact that all characters vanish at $z_+ = z_-$ is related to
the existence of the simple free field theory $\CS$ with $\cag$ symmetry,
defined in equation \cthree.
It has $k^+ = k^- = 1$ and hence $c=3$. 
Note that the free $\CS$-theory defined by $U, Q^{A\dot A}$
is always present as a subsector of any theory with $\cag$ symmetry.
One can always make a ``GKO coset construction'' factoring out the
$\CS$ theory and leaving a $W$-algebra type symmetry $\tl \cag$.
The characters of the quotient $\tilde \cag$
$W$-algebra are nonvanishing for short representations, and have a first
order zero for long representations.


\subsec{Constraints on the spectrum}

In general we expect that $\CH_{RR}$ is completely decomposable
in irreducible $\cag_{\rm left}\oplus \cag_{\rm right} $ representations, i.e.
\eqn\fulldec{
\CH_{RR} = \bigoplus_{\rho,\tl \rho} N(\rho,\tl \rho) \rho \otimes \tl \rho
}
with nonnegative integer degeneracies $N(\rho,\tl \rho) $.
We now describe what the index can teach us about the degeneracies
$N(\rho,\tl \rho) $.

The left-index is an expression of the form:
\eqn\ione{
I_1 = \sum_{\mu = 1 }^{k-1} \sum_{\Gamma}
  \Theta_{\mu,k}^-(\omega,\tau) 
\overline{\Xi_{\mu,u,\tl u}( \tau, \tilde\omega_+,
\tilde \omega_-)}  q^{u^2/k}\bar q^{\tl u^2/k}
}
where $\Gamma= \{(u,\tl u)\}$ 
is the set of $U(1)\times \widetilde{U(1)}$ charges.
Now, $\overline{\Xi_\mu}$ has contributions from both
the short and long representations:
\eqn\decopms{\eqalign{
\Xi_{\mu,u,\tl u}
= & \sum_{{\l^+ + \l^- = (\mu+1)/2 \atop \tl \l^+, \tl \l^-}}
(-1)^{2\l^-+1} n(\l^+,\l^-,u;\tl \l^+ , \tl \l^-, \tl u)
~\SCh_{(\tl \l^+,\tl \l^-)}(\tau, \tilde\omega_+, \tilde \omega_-) + \cr
& + \sum_{{\l^+ + \l^- = (\mu+1)/2  \atop \tl \rho: {\rm long}}}
(-1)^{2\l^-+1} N (\l^+,\l^-,u;\tl \rho)
~\SCh_{\tl \rho} (\tau, \tilde\omega_+, \tilde \omega_-)
}}
where the first line contains the sum over
short representations on both left and right,
and the second line contains the sum over $(short,long)$ representations
(with fixed $U(1)$ charge $\tilde u$).
For convenience, we denoted the degeneracies of the short representations
by $n(\l^{\pm},u;\tl \l^{\pm}, \tl u):=N(\l^{\pm},u;\tl \l^{\pm}, \tl u)$.
Similarly, the left-right index $I_2$ knows only about the short
representations. Specifically, it has the form
\eqn\indxtwofr{
I_2 = \sum_{\mu,\tl \mu=1}^{k-1}\sum_{\Gamma}
 D(\mu, u; \tl \mu,\tl u ) \Theta_{\mu,k} \,\overline{\Theta_{\tl \mu,k} } 
\,q^{u^2/k}\,\bar q^{\tl u^2/k}
}
with
\eqn\degnes{
D(\mu, u; \tl \mu,\tl u )
= \sum n(\l^{\pm},u; \tl \l^{\pm},\tl u)(-1)^{2\l^- + 2\tl \l^-}
\delta_{\mu, 2(\l^+ + \l^-)-1, ~\mod~ 2k}
\,\delta_{\tl \mu, 2(\tl \l^+ + \tl \l^-)-1, ~\mod~ 2k}
}
The multiplicities of the representations that occur in
the decomposition of $I_1$ and $I_2$ 
can be both positive and negative integers.
These multiplicities are related to $N(\rho, \tl \rho) \ge 0$
via \decopms\ and \degnes, respectively.


\subsec{Constraints of modular invariance}

The partition function of the theory in the odd 
spin structure is the supercharacter.
This should be ``modular invariant'' in the sense that it satisfies
 the transformation rule:
\eqn\modinvce{
Z_{RR}\Bigl( {a\tau+b\over c \tau + d} , 
	{\omega_+\over c \tau + d} , {\omega_-\over c \tau + d}; \dots\Bigr)
= e^{\Phi} Z_{RR}(\tau, \omega_+,\omega_-; 
\bar \tau, \tl \omega_+,\tl \omega_-)
}
where
\eqn\phasephi{
\Phi = 2\pi i \Biggl( k^+ {c \omega_+^2\over c\tau + d} 
+ k^- {c \omega_-^2\over c\tau + d}
\Biggr) - 2\pi i \Biggl( k^+ {c \tl \omega_+^2\over c\bar \tau + d} + k^- {c \tl\omega_-^2\over c\bar \tau + d} \Biggr)
}
The phase comes from the singular term in the $JJ $ ope. 
The transformation rule
\modinvce\ shows that $Z_{RR}$ is a  Jacobi form of weight $(0;0)$ and
index $(k,k; k,k)$ \ez.

The modular transformation rule \modinvce\ translates into a 
modular property of  the
index $I_2$:
%
%
\eqn\modtrmnitwo{ I_2(  {a\tau+b\over c \tau + d} , {\omega\over c
\tau + d} ,{\tl \omega\over c \bar \tau + d}) = (c \tau + d)(c
\bar \tau + d) e^{\Phi} I_2(\tau, \omega, \tl \omega) }
where now the phase is
\eqn\newfphs{
\Phi = 2\pi i \Bigl( k {c \omega^2\over c\tau + d}
\Bigr) - 2\pi i \Bigl( k {c \tl \omega^2\over c\bar \tau + d} \Bigr)
}

We now show how the modular covariance \modtrmnitwo\
of the index $I_2$ together with the simple 
transformation properties of level $k$ theta functions
can   be used to constrain the spectrum. As a simple example
suppose that   the degeneracies have the property:
\eqn\decouple{
n(\l^+,\l^-,u; \tl \l^+,\tl \l^-,\tl u) = n(\l^+,\l^-; \tl \l^+,\tl \l^-) n(u,\tl u)
}
and suppose moreover   that $n(u,\tl u)$ is such that
\eqn\nranis{
Z_{\Gamma} = \sum_{u,\tl u} n(u,\tl u) q^{u^2/k} \,\bar q^{\tl u^2/k}
}
is a Siegel-Narain theta function, i.e. transforms as a  modular form of weight $(1/2,1/2)$.
In this case, using the modular transformation rules 
for level $k$ theta functions in Appendix D, we see that
\modtrmnitwo\  implies that
\eqn\modinvt{
\sum_{\mu,\tl \mu=1}^{k-1} D(\mu,\tl \mu) \chi^{(k-2)}_{(\mu-1)/2} 
\,\overline{\chi^{(k-2)}_{(\tl \mu-1)/2} }
}
must define a modular invariant
partition function for the bosonic $SU(2)$ WZW model at level $k-2$.
We use $\chi^{(r)}_{j}$ to denote the character of the affine level $r$
$SU(2)$ representation with spin $j$, thus
\eqn\kmchr{
\chi^{(r)}_j(\omega, \tau) = 
{ \Theta_{2j+1, r+2}^-(\omega, \tau)\over \Theta^-_{1,2}(\omega,\tau) }.
}

The constraint \modinvt\ on the
spectrum of theories with $\cag$ symmetry   
has been found before in ref. \OoguriDS.
This constraint relies crucially on \decouple. As
we will see in the example considered below, the constraint \decouple\
on the spectrum  doesn't apply
to certain cases of interest. Note that  for a modular form
of nonzero weight  (such as $Z_{\Gamma}$)
the constant term  $\sim q^0 \bar q^0$  is {\it not}  modular invariant,
so in theories where \decouple\ does not hold the above
constraints do not apply to the spins even in the charge zero sector.

A more general class of theories for which there are simple constraints on
the spectrum from modular invariance  can be easily described by
extending the definition \defnewzee\ to include
an extra operator keeping track of the $U(1)$  charges of the states.
Thus we define:
\eqn\chargdfnze{
Z(\cdots, \chi_L,\chi_R):= e^{{k\pi \over 4 \tau_2} (\chi_L^2+\chi_R^2)}
{\Tr}_{\CH_{RR}} \biggl(\cdots e^{2\pi (\chi_L U_0 -\chi_R \tl U_0) } \biggr)
}
The exponential prefactor is included so that 
$Z$ has nice modular transformation
properties (i.e. without the phase in \modinvce), provided we transform
$$
(\chi_L, \chi_R) \to \bigl({\chi_L\over c \tau + d} ,
{\chi_R \over c \bar \tau + d} \bigr).
$$
{}From \chargdfnze\ we obtain an index $I_2(\CC, \chi)$.
Suppose that the $U(1)$ spectrum is such that we have
\eqn\itwochrg{
I_2(\CC,\chi) = \sum_{\beta \in \Lambda^*/\Lambda}
I_2^\beta  \Theta_\Lambda(\tau,0,\beta;P;\sqrt{k/2} \chi )
}
where $I_2^\beta$ are $\chi$-independent, 
and $\Theta_\Lambda$ is a Siegel-Narain
theta function for a   
lattice $\Lambda \cong \sqrt{k/2} II^{1,1}$. (See Appendix E
for our conventions on Siegel-Narain theta functions.) 
A nontrivial example of
such a theory may be found in sec. 4.6 below.
If $k$ is even, $\Theta_\Lambda$
transforms in a simple representation of the modular group. On the other hand
$I_2^\beta$  can be written  in terms of level $k$
theta functions of $\omega,\tl \omega$,
which also transform simply and we obtain a simple set of constraints on the
spectrum. 
The main example computed below, that of ${\rm Sym}^N(\CS)$, is in this
class of theories.

\subsubsec{\it Gauging the $U(1)$}

Given a theory $\CC$ with $\cag$ symmetry one can in principle
gauge a $U(1)$ subalgebra
to produce a theory $\hat \CC$ with $\tl \cag$ symmetry. 
(This is quite similar to the way
one derives the parafermion theories from the $SU(2)$ WZW model.)
Conversely, given a theory $\hat \CC$
with $\tl \cag$ symmetry we one can 
always tensor with the basic $\cag$ theory,  $\CS$,
to produce a new theory, $\check \CC$  with $\cag$ symmetry. We thus
have a transformation of theories $\CC \to \check \CC$ with $\cag$ symmetry.

We will now show that it is possible in some cases 
to describe how the index $I_2(\CC)$
is related to that of $I_2(\check \CC)$. 
We will limit ourselves to theories which
are rational with respect to $\cag$, and which satisfy the criterion \itwochrg.

First, let us describe the gauging process. 
We can gauge the axial $U(z) + \tl U(\zb)$ or
the vector $U(z) - \tl U(\zb)$ symmetry. For definiteness, let us gauge the
vector symmetry. Then gauging the symmetry projects 
onto vectors $\psi \in \CH(\CC)$
satisfying:
\eqn\project{
\eqalign{
U_n \psi & = 0 \qquad\qquad n > 0 \cr
\tl U_n \psi & = 0 \qquad\qquad n > 0 \cr
(U_0 - \tl U_0) \psi & = 0 \qquad\qquad  \cr}
}
The subspace satisfying \project\ is a representation of $\tl \cag$. However,
from the decomposition of characters $\SCh^{\cag} = \SCh^{\tl \cag} \SCh^{\CS}$
it is clear that the theory is infinitely degenerate with respect to $\tl \cag$,
and we wish to have a theory which is   rational with respect to $\tl \cag$.
Therefore we must also impose
\eqn\extra{
Q_r^{A\dot B} \psi =0  \qquad\qquad r>0
}
This theory is still infinitely degenerate because there will be infinitely
many primaries with $u_L =u_R$. However, if we consider the collection of
charge vectors:
\eqn\chrglatt{
\bigl\{ \sqrt{\textstyle{ k\over 2}}(u_L; u_R) \bigr\} \subset \IR^{1,1}
}
where $(p_L;p_R)\in \IR^{1,1}$ has metric $p_L^2 - p_R^2$, then we can
consider the case where these values lie in a lattice $\Lambda^*$, where
$\Lambda$ is an even integral lattice isomorphic to $\sqrt{k\over 2} II^{1,1}$.
Now, we can identify states whose charge 
vectors differ by a vector in $\Lambda$.
In this way we produce the gauged theory $ \hat \CC$ where $\hat \CC$
which is a rational   $\tl \cag$ theory.

The relation of the index of $\CC$ to that of $\hat \CC \otimes \CS$ is easily
stated. We decompose $I_2$ according to \itwochrg. Then
\eqn\nesxin{
I_2(\hat \CC \otimes \CS) = 
\biggl( \sum_{\beta_L = \beta_R} I_2^\beta\biggr) I_2(\CS)
}
One can check that the right hand side of \nesxin\ has the correct
modular transformation properties using the transformation properties of
Siegel-Narain theta functions of higher level.


\subsec{Relation to the $\CN=2$ elliptic genus}

In $\CN=2$ theory the procedure analogous to the operation
\soperat\
is to evaluate the character
\eqn\ntwoindx{
{\Tr}_R \bigl[q^{L_0-c/24} z^{J_0}\bigr]
}
and set $z=-1$. When applied to the left-moving sector
this defines the elliptic genus of the theory.%
\foot{Thus, the contribution of the right-movers gives a holomorphic
function of $\bar q, \tl z$. Throughout this section we suppress
mention of the right-moving sector for simplicity.}
In this case, only the states with $L_0=c/24$
can contribute, and hence the index is an integer. It counts the
R ground states. Similar remarks hold for the
small $\CN=4$ algebra.   In our case,
because of the larger $\cag$ symmetry we are able to keep
$z$ generic and yet maintain independence under deformation of
parameters. The states which are
contributing are not BPS with respect to the $\CN=2$
subalgebras, but they are nevertheless protected by the large $\CN=4$ symmetry.
That is why the $\cag$-index is valued 
in the ring of level $k$ theta functions,
rather than integers.

Note that, in striking contrast to $\CN=2$ and small $\CN=4$
theories $h-c/24$ is typically {\it positive} for Ramond
groundstates in theories with $\cag$ symmetry. Indeed, note that
the BPS bound in the R sector can be written $h-c/24 \geq
\mu(\rho)^2/4k > 0$. This has interesting implications, discussed
below, for the geometrical interpretation of the index.

\def\dm{{\dot -}}

The $\cag$ algebra has $\CN=2$ subalgebras. These are studied
in \GunaydinZW, so it is interesting to ask about how the
index is related to the elliptic genus with respect to these
$\CN=2$ subalgebras.  Up to an $SU(2)\times SU(2)$ transformation we can
 consider the generators
\eqn\ntwogens{
\eqalign{
\CG^+_m & = i \sqrt{2} G_m^{\dot +,+} \cr
\CG^-_m & = i \sqrt{2} G_m^{\dot -,-} \cr
J_m & = 2i \bigl(\gamma A^{+,3}_{m} - (1-\gamma) A^{-,3}_n\bigr) \cr}
}
Together with $L_n$, these generate an $\CN=2$ subalgebra of $\cag$ with
\eqn\subsl{
\eqalign{
\{ \CG^{+}_m, \CG^-_n\} & = 
2L_{n+m} + (m-n) J_{m+n} + {c\over 12} \delta_{n+m,0} (4m^2-1) \cr
[J_m, \CG^+_n] & = - \CG^+_{n+m} \cr
[J_m, \CG^-_n] & = + \CG^+_{n+m} \cr}
}

If $\Phi^{m^+, m^-}$ is an NS-sector  BPS multiplet of $\cag$
with $kh = k^+ \l^- + k^- \l^+ + (\l^+ - \l^-)^2$ then a computation
of the charge under $J_0$ shows that we do not get chiral primaries
under the $\CN=2$ subalgebra unless $\l^+ = \l^- = \l$ in which case
$\Phi^{\l,-\l}$ is anti-chiral primary and $\Phi^{-\l,\l}$ is chiral primary.

It is interesting to consider spectral flow using this $\CN=2$ subalgebra.
$Q_n^{{\dot +},+}, Q_n^{\dm,-}$ have charges $-1,+1$ while
\eqn\suneutl{
\eqalign{
[J_m, Q_n^{{\dot +},-} ] & = (1-2\gamma) Q^{{\dot +},-}_{n+m} \cr
[J_m, Q_n^{\dm,+}  ] & = -(1-2\gamma) Q^{\dm,+}_{n+m} \cr}
}
Note that $\gamma=\half$ is an interesting special case. Then spectral
flow by 1 unit takes the NS sector to the R sector, but \suneutl\
remain in the NS sector. Thus, one must be careful about referring to
$NS$ and $R$ sectors since they can mean different things for the $\cag$
algebra and the $\CN=2$ subalgebra.

It might appear that one can introduce a new index using this $\CN=2$
subalgebra. In fact, the index turns out to be already encoded in
the index we have already introduced. To be more precise, let us
suppose that the $NS$ sector coincides for the $\cag$ and $\CN=2$ algebra.
The $\CN=2$ index can be expressed in terms of the $NS$-sector trace
\eqn\nssect{
{\Tr}_{\CH_{NS}}\bigl[ e^{2\pi i \tau (L_0 - \half J_0) } e^{i \pi J_0}\bigr]
}
Now using $\cag$ spectral flow we find that \nssect\ is, up to a
simple prefactor  the trace \defchrs\ evaluated at
\eqn\frels{
\omega_+ = \omega_- = \half (\tau+1) (1-\gamma)
}
We may conclude two things from this: First, the $\CN=2$ index in fact
vanishes, since $R$-sector $\cag$ characters always vanish when
$\omega_+ = \omega_-$. Second, the index \soperat\ is a generalization
of the ``new index'' ${\Tr} q^{L_0-c/24} F (-1)^F$ of
\CecottiQH.


\subsec{An index for the Wigner contraction $\CA_{k^+,\infty} $ }

A Wigner contraction of $\CA_{k^+,k^-}$ produces an extension of a
small $\CN=4$ algebra.
This contraction is relevant to the study of holography for strings on
$AdS_3 \times \S^3 \times T^4$. In this context ref.
 \MaldacenaBP\ introduced an index for computing BPS states. In this section
we clarify the relation of the index of \MaldacenaBP\ to the present index.

We define the contraction by   separating
the zeromodes of the $SU(2)^-$ current algebra and defining
\eqn\wigner{
\eqalign{
\tilde A^{-,i}_n &  = {1\over \sqrt{k^-}} A^{-,i}_n  \qquad n\not=0 \cr
\tilde A^{-,i}_0 &  = A^{-,i}_0 , \cr
 \tilde U & = {1\over \sqrt{k^-}} U, \cr
 \tilde Q^a  & = {1\over \sqrt{k^-}} Q^a\cr}
}
We then take the limit $k^- \to \infty$ in the commutation relations, holding
$\tilde A^{-,i}, \tilde U, \tilde Q^a$ fixed to obtain the small $\CN=4$
algebra extended by a $U(1)^4$ current algebra {\it together with} a
global, finite-dimensional  $SU(2)^-$ algebra spanned by $A_0^{-,i}$.
The $U(1)^4$ currents form a ${\bf 3} \oplus {\bf 1}$ and the
fermionic partners are in the ${\bf 2} \oplus {\bf 2} $. If desired, one
could make a further contraction of the global $SU(2)^-$ algebra to
$U(1)^3$, thus, supplying the remaining zeromodes of the $U(1)^4$
current algebra.

The key remark is the observation of \refs{\PTone,\PTtwo}
that the characters for $\cag$ have well-defined limits
for $k^- \to \infty$, holding all other quantum numbers fixed.
By inspection one finds that the  massless characters have the form:
\eqn\masslss{
\SCh^{\cag,R}_0 = (z_+ + z_+^{-1} - z_- - z_{-}^{-1}) G_0(k^+,k^-,q,z_+,z_-)
}
while the massive characters have the form
\eqn\msvve{
\SCh^{\cag,R}_m = (z_+ + z_+^{-1} - z_- - z_{-}^{-1})^2 G_m(k^+,k^-,q,z_+,z_-)
}
where the functions $G_0$ and $G_m$ have the following properties:
First, these functions have smooth limits for $k^- \to \infty$. Second,
the functions are nonzero at $z_+= z_-$ and do not get extra zeroes
either at $z_+ = z_- = 1$ or as $k^- \to \infty$.

Now working  in the limit $k^-=\infty$, ref.  \MaldacenaBP\ sets $z_-=1$.
Notice that in this case the prefactor
\eqn\degn{
(z_+ + z_+^{-1} - z_- - z_{-}^{-1}) = z_+^{-1}( z_+-1)^2
}
develops a second order zero at $z_+=1$. From \masslss\msvve\ it is clear that
to form a nonzero index one must take 
two derivatives with respect to   $z_+$ and
set $z_+=1$. This is the index of \MaldacenaBP.
%
%
The precise relation between the indices  is that
\eqn\mssindxi{
{d\over dz}\Bigl\vert_{z=1} I_{1}(\CC)
}
is, up to a simple numerical factor, identical with the index of \MaldacenaBP.
At finite $k^-$ the RHS of \mssindxi\ 
is a function of $q,\bar q, \tilde z_+, \tilde z_-$.
As $k^- \to \infty$ the $q$-dependence drops 
out and the expression becomes a holomorphic
function of $\bar q, \tilde z_+$ at $\tilde z_-=1$.

In hindsight, the index of \MaldacenaBP\  was unnecessarily
restrictive. Since the small $\CN=4$ algebra has a global $SU(2)^-$ symmetry
one could have kept  $z_- \not=1$, and used the index described in the
present paper. Thus, it  might be of 
interest to reexamine the partition function in the
$AdS_3 \times \S^3 \times T^4$ background using the more powerful index
$I_1$. A word of warning is in order here. In the limit $k^- \to \infty$,
we have $\S^3 \times \S^1 \to \IR^3 \times \S^1$.
The zeromode algebra $A_0^{-,i},U_0$ is broken by the identifications
one would want to make to produce $T^4$. Note, in this connection,
that the argument of \MaldacenaBP\ for invariance of the index
fails precisely for states with nonzero $U(1)$ charges (see remarks under
 eq. 3.11 of \MaldacenaBP). Thus, the index of \MaldacenaBP\ only
applies to states where we could replace $T^4 \to \IR^4$, and this
coincides with the $k^- \to \infty$ limit of $\S^3 \times \S^1$ in
the charge zero sector.

\subsec{Geometrical interpretation of the index}

In this section we would like to call attention to some interesting open
problems concerning the
geometrical interpretation of the indices $I_1,I_2$
defined above in the case of a supersymmetric
nonlinear sigma model with
target space $X$ having $\cag_{\rm left} \oplus \cag_{\rm right} $
symmetry.

In $\CN=2$ sigma models, with a Kahler target space $X$,
the elliptic genus $\chi(q,y)$ computes topological invariants
of the space. The $q\to 0$ limit gives the character \KawaiJK:
\eqn\qzero{
\sum_{r,s=0}^{\dim X}  (-1)^{r+s} h^{r,s}(X) y^{s-\half\dim _c X}
}
The same is true for small $\CN=4$ sigma models with hyperkahler
target space.

Moreover, if one does not take the $q\to 0 $ limit then the
elliptic genus $\chi(q,y)$ has the interpretation of
a.) computing an infinite set of indices of Dirac operators
coupled to bundles on $X$ and b.) more conceptually, is the
$U(1)$-equivariant index of the Dirac operator on the loop
space of $X$ \refs{\SchellekensYI,\SchellekensYJ,\PilchEN,\witteneg,
\AlvarezWG,\AlvarezDE,\WindeyAR,\PilchGS,\LercheQK,\LercheCA,\KawaiJK}.

It might be very interesting to generalize these statements
to the large $\CN=4$ indices $I_1,I_2$.
The reason is that, as we have remarked above
  $h-c/24$ is positive
in the R sector. From the geometrical point of view
the   reason for this is   that
for fixed $k^\pm$ the ``size'' of the target space is fixed.
That is, one cannot, at {\it fixed} $k^\pm$, take an $\alpha' \to 0$ limit
of the target space and recover classical geometry. For this reason,
the index for $\cag$ theories is ``more stringy'' than its $\CN=2$
and small $\CN=4$ counterparts. This is closely related to the
fact that the index is a theta function, rather than an integer.

In order to address this problem it is necessary to clarify
the geometrical conditions on a $\sigma$-model such that it has
$\cag$ symmetry. 
References \refs{\deBoerRH,\OpfermannQD} have some interesting relevant
material  for this problem, but the precise statement does not appear to be
available. Next one would wish to find a
geometrical interpretation of the BPS condition \GPTVP
\eqn\bpscond{
(\tilde G_{-\half})^{(A_1}_{(\dot A_1} 
\Phi^{A_2\cdots A_n) }_{\dot A_2\cdots \dot A_m)} =0
}
in terms of differential equations on some tensors on the target space.


\newsec{Application to symmetric product theories }

The elliptic genus of symmetric product theories
 has proven to be quite useful in investigations of
the AdS/CFT correspondence in the case of two-dimensional
conformal field theories \refs{\StromingerSH,\deBoerIP,
\MaldacenaBP,\DijkgraafFQ,\MooreFG}. It is therefore natural to
devote some special attention to symmetric product theories
with $\cag$ symmetry.

Suppose a CFT $\CC_0$ has $\CA_{\gamma}$ symmetry.
We would like to define a new theory, the
symmetric product orbifold, with   $\CA_{\gamma}$ symmetry.
We define the
theory by applying the symmetric product construction to $\CH_{RR}$,
and then construct the remaining sectors by spectral flow.
The Hilbert space in the RR sector has the form:
\eqn\symmprdrr{
\CH_{RR}({\rm Sym}^N(\CC_0)) = \bigoplus_{(n)^{\l_n} } \bigotimes_n
{\rm Sym}^{\l_n}(\CH_{RR}^{(n)}(\CC_0))
}
where $\CH_{RR}^{(n)}(\CC_0)$ is the twisted sector corresponding
to a cycle of length $n$ and  we sum over partitions $\sum n \ell_n = N$,
in the standard way.
In order for the  Hilbert space \symmprdrr\ to  be a representation of
the R-moded $\CA_{\gamma}$ algebra it is quite important that we take
a {\it graded} tensor product, relative to the $\IZ_2$-grading of
$(-1)^F$. With this understood \symmprdrr\   can be used to construct
an $\CA_{\gamma}$-invariant theory.

In \DijkgraafXW\ a procedure was given for expressing the partition function
of symmetric product theories in terms of that of the parent theory.
The argument is reviewed briefly in appendix F.
 There are two ways of stating the result. One involves infinite products and
the other involves Hecke operators.
For purposes of computing the index it is more useful to give the
 formulation in terms of Hecke operators.
We introduce the generating functional:
\eqn\generatfun{ \CZ := 1+ \sum_{N\geq 1} p^N {\STr}_{{\rm
Sym}^N(\CC_0) } \Bigl[q^{L_0-c/24}\tilde q^{\tilde L_0-c/24}
z_+^{2A_0^{+,3}} z_-^{2A_0^{-,3}} \tilde z_+^{2\tilde A_0^{+,3}}
\tilde z_-^{2\tilde A_0^{-,3}} \Bigr]}
Then the result of appendix F is:
\eqn\heckeform{
\log \CZ = \sum_{M=1}^\infty p^M T_M Z_0
}
where   $Z_0$ is the partition function of 
$\CC_0$ and the Hecke operator is defined by
\eqn\teenn{
T_M Z_0 := {1\over M}  \sum_{M = ad} \sum_{b=0}^{d-1} 
Z_0\Bigl({a \tau + b \over d},
a\omega_+, a \omega_-  ; {a \bar \tau + b \over d} ,
a\tl\omega_+, a \tl \omega_- \Bigr)
}
The first sum is over all factorizations of 
$M$ into a product of integers $a \times d$.
%
%
%

Now let us compute the index.  The coefficient $Z_N$ of $p^N$ in
\generatfun\ will be a polynomial in the Hecke transforms $T_M Z_0$:
\eqn\polyexp{
Z_N = T_N Z_0 + \cdots + {1\over N!} (T_1 Z_0)^N
}
All summands in \polyexp\ beyond the   leading term are products
of more than one Hecke transform.
Since   $T_MZ=0$ for $z_-=z_+$ the only term which survives the
index operation
is the linear term. Thus
 the index simplifies enormously compared to the full partition function
and we have
\eqn\indxsymprd{
I_1({\rm Sym}^N(\CC_0)) = - z_+ {d\over dz_-}  T_NZ_0
}
evaluated at $z_- = z_+=z$.

\subsec{Interpreting the states which contribute to the index}

We would like to make three remarks on the interpretation of various
terms in the index.

First, we make a distinction between 
{\it short-string} and {\it long-string } contributions.
The Hecke operator $T_NZ_0$ always involves two kinds 
of terms, $a=N, d=1$ and $a=1, d=N$.
(When $N$ is prime these are the only contributions,
but in general there are other contributions.)
If we view the trace as the partition function 
of the conformal field theory on a
torus, then these  two contributions have natural interpretations 
in terms of covering tori.
The $a=N,d=1$ term corresponds to strings 
which are ``long strings'' in the Euclidean
time direction. The $a=1, d=N$ term 
corresponds to strings which are ``long strings''
in the space direction. In general, 
when $N$ is not prime we have a combination of
strings which are $d$-long in the space 
direction, wrapping $N/d=a$ times in the Euclidean
time direction.
 In the supergravity interpretation of the holographic dual to
${\rm Sym}^N(\CC_0)$ it turns out that the $a=N,d=1$ term
corresponds to the contribution of supergravity particles
to the index while the  $a=1,d=N$ term presumably comes from
other geometries, possibly conical defect geometries 
\LuninJY.

Secondly, we would like to discuss  
the contributions of specified twisted  sectors in terms of
Hecke operators.
In order to associate contributions of particular states with
various terms in the index we
introduce formal variables $x_n$ to keep track of cycles of length $n$ and
we  multiply ${\rm Sym}^{\ell_n}(\CH^{(n)}(\CC_0))$ by $x_n^{\ell_n}$.
If one follows through the derivation in appendix $F$ one discovers that
we have the modified Hecke operator:
\eqn\teennmod{
\hat T_M Z_0 := {1\over M}  \sum_{M = ad} \sum_{b=0}^{d-1} (x_d)^a 
Z_0\Bigl({a \tau + b \over d},
a\omega_+, a \omega_-  ; {a \bar \tau + b \over d} ,
a\tl\omega_+, a \tl \omega_- \Bigr)
}
Using this expression in \polyexp\ and collecting terms proportional
to a given monomial  $\prod_n x_n^{\ell_n}$ gives an expression for the
contribution of that twisted sector to the full partition function.
It  is then clear that the   contribution 
to \indxsymprd\  from the $N=ad$ term in
\teenn\ comes entirely from states which are  in the twisted sector:
\eqn\contribut{
{\rm Sym}^{a}\bigl(\CH^{(d)}(\CC_0)\bigr)
}
in the sum \symmprdrr.

Finally, we would like to mention the {\it inheritance principle.}
This is a useful observation for  interpreting which states contribute to the
index. Recall from \condi\ - \condii\ that the states which contribute to the
index satisfy
\eqn\inherit{ k(L_0 - c/24)  =   U^2 + (A_0^{+,3} + A_0^{-,3})^2 }
If there is a state in $\CC_0$ which saturates the bound, then the
corresponding state in the long string sector of ${\rm
Sym}^N(\CC_0)$ will also saturate the bound, since $k \to Nk$
while  the eigenvalue $\Delta$ of $L_0 - c/24 $ goes to $\Delta/N$
and the eigenvalues of $A_0^{+,3}$ and $ A_0^{-,3}$ remain
unchanged.

\subsec{The Hecke algebra}

The Hecke operators defined in \teenn\ satisfy a beautiful
algebra.\foot{For some material on  Hecke theory for Jacobi
forms see the book of Eichler and Zagier \ez.}
In order to describe this algebra, let us consider
operators acting on the space of functions $f(\tau,\omega)$
such that $f(\tau+1,\omega) = f(\tau,\omega)$.
For any positive integer $n$ let us introduce the operators
\eqn\defhops{
\eqalign{
(U_n f)(\tau,\omega) & := {1\over n} \sum_{b=0}^{n-1} 
f\bigl({\tau+b \over n}, \omega\bigr) \cr
(V_n f)(\tau,\omega) & := {1\over n} f(n \tau, n \omega) \cr
(W_n f)(\tau,\omega) & := {1\over n} f(\tau, n \omega) \cr}
}

Note that we obviously have
\eqn\fisrtresl{
\eqalign{
V_{n_1} V_{n_2} & = V_{n_2} V_{n_1} = V_{n_1 n_2} \cr
W_{n_1} W_{n_2} & = W_{n_2} W_{n_1} = W_{n_1 n_2} \cr
V_{n_1} W_{n_2} & = W_{n_2} V_{n_1}   \cr}
}
Slightly less obvious is
\eqn\secresl{
U_{n_1} U_{n_2} = U_{n_2} U_{n_1} = U_{n_1 n_2}
}
but this follows from noting that if $b_i \in \{ 0, \dots , n_i-1\}$ then
$(b_1 + n_1 b_2) ~\mod~ n_1 n_2$ 
takes all values in $\{ 0, \dots , n_1 n_2 -1\}$.

The algebra satisfied by $U_{n}$ and $V_{n}$ is a little more intricate.
First, if $(n_1, n_2)=1$ then
\eqn\thrrel{
V_{n_1} U_{n_2} = U_{n_2} V_{n_1}
}
(this uses $f(\tau+1,\omega) = f(\tau,\omega)$). On the other hand,
$V_{n}$ and $U_{n}$ do {\it not} commute. More generally, $V_{n_1}$ and
$U_{n_2}$ do not commute if $(n_1,n_2)>1$. An easy computation shows
\eqn\frht{
U_{n} V_{n} = {1\over n} W_{n}
}
while
\eqn\frhti{
(V_{n} U_{n} f)(\tau,\omega)
= {1\over n^2} \sum_{b=0}^{n-1} f(\tau + {b\over n}, n \omega) .
}

Now, we can write the Hecke operators $T_{n}$ in terms of $U$ and $V$ as
\eqn\heckuv{
T_{n} = \sum_{d\vert N} V_a U_d
}
where $a := N/d$, as usual. Note that it follows immediately
from the above relations on $U,V$ that if $(n_1, n_2)=1$ then
\eqn\ahecksi{
T_{n_1} T_{n_2} = T_{n_2} T_{n_1} = T_{n_1 n_2}.}
Now, let $p$ be a prime number. A
simple computation using \frht\ shows that
\eqn\primpow{
T_{p} T_{p^r} = T_{p^{r+1}}  + {1\over p } T_{p^{r-1}} W_{p} .
}
These relations are elegantly summarized in the formula
\eqn\genrati{
\sum_{r=0}^\infty T_{p^r} X^r = {1 \over 1- X T_{p} + X^2 {1\over p} W_{p} }
}
Now, since $T_p = U_p + V_p$ we have
\eqn\factorit{
1- X T_p + X^2 {1\over p} W_p = (1- XU_p)(1-X V_p)
}
Thus, putting $X= p^{-s}$,
we may write the elegant formula summarizing the Hecke operators
in terms of $U_p,V_p$:
\eqn\eulerprod{
\sum N^{-s} T_N = \prod_{p ~ {\rm prime} } 
\biggl( {1\over 1- p^{-s} V_p} \biggr)
\prod_{p ~ {\rm prime} } \biggl( {1\over 1- p^{-s} U_p} \biggr)
}

Finally, we would like to mention a physical interpretation of the
various operators we have just discussed.
The operators $U_n$ and $V_n$ have the physical interpretation
of ``creating'' the long strings and short strings of the symmetric product,
respectively. Thus, \heckuv\ has the interpretation that $U_d$
``creates'' a long string background and $V_a$ then ``adds'' short
string excitations to it.
If we take an {\it ensemble}  of theories $\syms$ for all $N$
(as in Matrix theory) then the summation on the LHS of
\eulerprod\ might have a physical meaning. Then the Euler product formula
might be very interesting from a physical point of view.

\subsec{Multiple symmetric products}

An interesting variation on the symmetric product construction is that
of multiple symmetric products ${\rm Sym}^N {\rm Sym}^M(\CC_0)$. This
may be thought of as an orbifold of $\CC_0^{NM}$ by the
{\it wreath product} $S_N \wr S_M$, which is a proper subgroup of
$S_{NM}$.

The wreath product is best thought of by thinking of a set of
$NM$ objects partitioned into $N$ sets of $M$ objects each, e.g.
points $X^{i a}$ with $1\leq i \leq M$ and $1\leq a \leq N$.
We are allowed to permute the $i$'s for a fixed $a=a_0$, holding
$X^{ia}$ fixed for $a\not= a_0$. We are also allowed to permute
$X^{ia} \to X^{i \sigma(a)}$.

The amusing point about this construction is that, for
$N,M$ relatively prime, the index $I_1$ is the {\it same }
as that for the symmetric product ${\rm Sym}^{NM}(\CC_0)$.
However, when $(N,M)>1$ the index is different, because of
\primpow. Some examples of this are given in Appendix G.
Thus ${\rm Sym}^{NM}(\CC_0)$ and ${\rm Sym}^N{\rm Sym}^M(\CC_0)$
are distinct theories. Indeed we can keep iterating to produce
a large variety of theories.

These theories offer an interesting cautionary tale in thinking
about the AdS/CFT correspondence, because they show that
there are {\it many} different CFT's with $\cag$ symmetry and
small gap above the ground state in the Ramond sector, and
hence the qualitative spectra of black holes cannot be used
to deduce that, say, the holographic dual of
$AdS_3 \times \S^3 \times \S^3 \times \S^1$ is ${\rm Sym}^N(\CS)$.


\newsec{The index for ${\rm Sym}^N(\CS)$}

In this section  we present a computation for a family
of theories with $\cag$ symmetry.
We will consider the index for the theory
$\CC={\rm Sym}^N(\CS)$. This theory has
$\cag$ symmetry with $k^+ = k^- = N$, hence $k=2N$, $c=3N$.

The main motivation for the computation we are going to present
is the search for a holographic dual for string theory
on $AdS_3 \times \S^3 \times \S^3 \times \S^1$. It was proposed in
\refs{\ElitzurMM,\deBoerRH} that under certain conditions on the fluxes
of RR and NSNS tensor fields, the holographic dual is on the moduli
space of the above theory. See \GukovYM\ for more details, and
for the applications of the computation of the present section.

\subsec{Summary of results for the index}

The formula for $I_2(\CC)$ is rather complex. To simplify matters,
we assume that $N$ is prime and we restrict attention to the
charge zero sector. (Generalizations for nonprime $N$ and nonzero
charge are in sections 5.6, 5.8  below.) The result for the
left-right index is
\eqn\itoosym{
I_2^0(\CC) = 
(N+1) \vert \Theta^-_{N,k}\vert^2 + \sum_{{1 \le \mu \le 2N-1 \atop \mu=1~\mod~ 2}}
\biggl\vert \Theta^-_{\mu,k} + \Theta^-_{2N-\mu,k}\biggr\vert^2
}
Here and below the conjugation operation implied in $\vert \Theta \vert^2$
 takes $\omega_\pm\to \tilde \omega_\pm $ and acts as complex conjugation.

As for the index $I_1(\CC)$, $\Xi_{\mu}$ defined in \ione\ has the form
$\Xi_{\mu}^{ss} + \Xi_{\mu}^{ls} $, corresponding to the contribution of
short strings and long strings.  We find that
\eqn\ximusg{
\Xi_{\mu}^{ss}= \delta_{\mu,N} \sum_{\l^-=1/2}^{N/2}
 (-1)^{2\l^--1} \SCh\Bigl({N+1\over 2}- \l^-, \l^-\Bigr)+ \cdots
}
where here and below $+\cdots$ indicates long representations.

Moreover we have
\eqn\expbhxi{
\Xi^{ls}_{\mu} = \SCh\Bigl({\mu+1\over 4},{\mu+1\over 4} \Bigr) +
\SCh\Bigl({N+1\over 2} -{\mu+1\over 4},{N+1\over 2} -{\mu+1\over 4}\Bigr) 
+ \cdots
}
for $1\leq \mu \leq 2N-1$, $\mu=1~\mod~ 4$,
and $\mu\not= N$,  while
\eqn\expbhxii{
\Xi^{ls}_{\mu} = -\SCh\Bigl({\mu-1\over 4},{\mu-1\over 4} \Bigr) -
\SCh\Bigl({N+1\over 2} -{\mu-1\over 4},{N+1\over 2} -{\mu-1\over 4}\Bigr) 
+ \cdots
}
for $3 \leq \mu \leq (2N-3)$, $\mu= 3 ~\mod~ 4$, and $\mu \not=N$.
If $\mu=N$ we have
\eqn\muennbh{
\Xi^{ls}_{\mu=N} = (-1)^{\half N (N-1)} 
\SCh\Bigl({N+1\over 4},{N+1\over 4} \Bigr) + \cdots
}

The simplest RR spectrum consistent with these results is
\eqn\simplest{
\eqalign{
& \bigoplus_{\l^-=1/2}^{N/2} \Bigl\vert \Bigl({N+1\over 2} - \l^-, \l^-\Bigr) 
\Bigr\vert^2 \cr
&
\bigoplus_{\l=1/2}^{(N-1)/4} 
\biggl\vert (\l,\l) + \bigl({N+1\over 2}-\l, {N+1\over 2}-\l\bigr)\biggr\vert^2
\oplus \biggl\vert  \Bigl({N+1\over 4} , {N+1\over 4} \Bigr) \biggr\vert^2 \cr}
}
where the first line comes from the short string and the
second from the long string contribution.  The short string states have
$h= {N\over 4}$ for all states, and the gap to the next excited state is
order $1$. The long string states $(a=1,d=N)$ have
\eqn\lngwt{
h= {N\over 8} + {(4\l -1)^2\over 8N}
}
and have small gaps $\sim 1/N$ to the first excited state.

While this is the simplest spectrum, it is not the unique spectrum consistent
with the index. The true spectrum might differ by a virtual representation
of index zero. 
These can always be expressed in terms of massive representations and
are expected to disappear upon turning on a 
generic perturbation -- barring some extra
symmetry protecting states.

Applying spectral flow to the representation \simplest\ gives:
\eqn\simplestns{
\eqalign{
& \bigoplus_{j=0}^{(N-1)/2} \vert (j,j)_{NS} \vert^2 \cr
&
\bigoplus_{j=0}^{(N-3)/4} 
\biggl\vert ({N-1\over 2}-j,j)_{NS} + (j, {N-1\over 2}-j)_{NS} \biggr\vert^2
\oplus \biggl\vert \Bigl({N-1\over 4} , {N-1\over 4} \Bigr)_{NS} 
\biggr\vert^2 \cr}
}
where the first line is from the short string contribution $a=N, d=1$
and the second from the long string  contribution $a=1,d=N$.
The short string states have
$h= j  $
while the long string states have:
\eqn\lngwt{
h= {N-1\over 4} + {(N-1-4j)^2\over 8N} .  }
In subsections 5.3 - 5.5  we give the proofs of these results.

\subsec{Interpreting states contributing to the index}

Before presenting the technical computation we
 give a direct construction of a set of states that accounts for
the nonvanishing index $I_2(\CC)$.

{\it Short string states} $(a=N,d=1)$: 
Here we are looking for $(bps,bps)$ states in
${\rm Sym}^N(\CH)$ where $\CH$ is the Hilbert space of $\CS$. Again we must
stress that this is a graded tensor product.  
The states are most simply described in the NS sector.
We consider states built out of the fermion operators $\psi_{-1/2}^a(i)$
(where $(i)$ refers to the factor in the symmetric product).
We may then introduce the operator
\eqn\ohab{
\CO^{ab} = \sum_i \psi_{-\half}^a(i) \tilde \psi_{-\half}^b(i)
}
Amongst the  states:
\eqn\nssctr{
 \CO^{a_1b_1} \cdots \CO^{a_nb_n} \vert 0 \rangle_{NS}
}
there is a subspace of states transforming as
 $SU(2)^4$ multiplets $({n\over 2},{n\over 2};{n\over 2},{n\over 2})$,
and having $h= n/2$. 
They are therefore BPS multiplets. Moreover, for $N>1$ they
are {\it not} descendents  in the BPS multiplet $\vert (0,0)\vert^2$
since the descendent fermion operator is 
$Q_{-\half}^a = \sum_{i=1}^N \psi_{-\half}^{a}(i) $.
Finally, the tower cuts off at $n=N-1$ by the Fermi statistics 
of $\psi_{-1/2}^a(i)$, i.e. for $n=N$ the state \nssctr\ is a descendent
$\vert Q_{-\half}^{a_1} \cdots Q_{-\half}^{a_N} \vert^2 \vert 0 \rangle$.

{\it Long string states}: $(a=1,d=N)$. Here we may use the
inheritance principle of \inherit. If we identify the states in the
$\CS$ theory which contribute to the index for $\CS$, then we automatically
find corresponding states in the long-string sector contributing to the
index.  We will see below that the index for $\CS$ comes from states
with $h-c/24 = \half m (m+1)$, $m\geq 0$ with spins of the form:
\eqn\intrptst{
\psi_{-1}^{+,+} \psi_{-2}^{+,+}\cdots \psi_{-m}^{+,+}\vert \Omega_\pm \rangle
}
where all the $SU(2)\times SU(2)$ spins are aligned, we have used
bispinor notation for the fermions and $\Omega_\pm$ are the two highest
weight states for the Ramond ground state of $\CS$. Now, applying the
inheritance principle we find a collection of multiplets contributing
$\vert (\ell,\ell)\vert^2 $  as in the
second line of \simplest.  By suitably flipping spins one
can obtain the off-diagonal representations in the same way
(\cf\ section 5.3 of \GukovYM).

Let us comment on BPS states which {\it do not} contribute to the index.
There is an algorithm by which one can,
in principle, determine the exact spectrum (not just BPS) 
of the theory $\CC$ using
the algebra of level $k$ theta functions. The reason is that $Z_0$
for $\CS$ can be expressed in terms of even level $1$ theta functions
$\Theta^+_{\mu,1}(\omega_+,\tau)$ and $\Theta^+_{\mu,1}(\omega_-,\tau)$.
The Hecke transform $T_M$ maps these to even level $M$ theta functions.
The even  theta functions form an algebra, and
the partition function of $\CC$ can be expressed
as a {\it linear} combination of  even level $N$ theta
functions of $\omega_\pm, \tl \omega_\pm $ 
(with modular forms as coefficients).
Finally, the space of affine $SU(2)$ characters $\chi^{(k)}_\mu$
is spanned by even level $k$ theta functions. Putting these facts together
gives an algorithm for determining the exact representation content of
the theory $\syms$.
We have not carried this out partly
because the expressions are complicated, and partly 
because it is only the index
which is expected to be an invariant on the moduli space of the
$\syms$ theory. Nevertheless, it is 
easy to identify some interesting BPS states which
come in cancelling pairs, and this we now describe.

One set of interesting BPS states comes in twisted sectors associated
to the conjugacy class $(n)(1)^{N-n}$ for $0<n<N$.
These states live in the summand
\eqn\twistste{
\CH^{(n)} \otimes {\rm Sym}^{N-n}(\CH)
}
of \symmprdrr.
In the RR sector the character will have a second order
zero at $z_+= z_-$  since it is the product
of two characters for $\cag$ at $(k^+,k^-)=(n,n)$ and at $(N-n,N-n)$
respectively.
Thus, we can expand in terms of massive characters.  Nevertheless
we should examine these sectors more closely since the 
massive characters are at threshhold.
In the NS sector of $\CH^{(n)}$, with $n$ odd,  one can construct
explicitly \GukovYM\   BPS states with
\eqn\twisfi{
h= \ell^+ = \ell^- = {(n-1)\over 4}
}
These give BPS states when combined with
certain other states in  ${\rm Sym}^{N-n}(\CH)$. One way to do this is
by
tensoring the BPS state   \twisfi\ with any of the four states:
\eqn\partdec{
\eqalign{
& \vert 0 \rangle_{{\rm left}} \otimes \vert 0 \rangle_{{\rm right}} \cr
& \biggl( \sum_{i=1}^{(N-n)} Q_{-1/2}^a(i)\biggr)
\vert 0 \rangle_{{\rm left}} \otimes \vert 0 \rangle_{{\rm right}} \cr
&  \vert 0 \rangle_{{\rm left}} \otimes \biggl(
\sum_{i=1}^{(N-n)} \tilde Q_{-1/2}^a(i) 
\vert 0 \rangle_{{\rm right}} \biggr) \cr
& \biggl( \sum_{i=1}^{(N-n)} Q_{-1/2}^a(i)\biggr) 
\vert 0 \rangle_{{\rm left}} \otimes
\biggl( \sum_{i=1}^{(N-n)} \tilde Q_{-1/2}^a(i) 
\vert 0 \rangle_{{\rm right}} \biggr) \cr}
}
If the spin $a$ is aligned with that of 
\twisfi\ then the resulting state will be BPS, but
for $n<N$ it is not a descendent.  Alternatively, we can consider
\eqn\altdpar{
\eqalign{
& \Phi \otimes \vert 0 \rangle_{N-n} \cr
&Q^{(n)}_{-\half} \Phi \otimes \vert 0 \rangle_{N-n} \cr
&\tl Q^{(n)}_{-\half} \Phi \otimes \vert 0 \rangle_{N-n} \cr
&Q^{(n)}_{-\half} \tl Q^{(n)}_{-\half}\Phi \otimes \vert 0 \rangle_{N-n} \cr}
}
These also form a cancelling quartet.  In the terminology of the AdS/CFT
correspondence \altdpar\ are ``single particle states'' while \partdec\ are
``multiparticle states.''

 A second way in which one can make new BPS states from the states \twisfi\ is
by tensoring (in the NS sector) with
the short string BPS states
in  ${\rm Sym}^{N-n}(\CH)$, which are   analogous to the
states \nssctr\ constructed above. The result is 
a set of BPS representations with
\eqn\setfo{
(h,\l^+,\l^-) = ({n-1\over 4} + {s\over 2} , 
{n-1\over 4} + {s\over 2}, {n-1\over 4} + {s\over 2})
\qquad  0 \leq s \leq N-n-1
}
on both left and right. Under spectral flow to the RR 
sector these states contribute
to the index $(-1)^{(n-1)/2 + s} \Theta_{N,k}$. For $N$ odd the sum over $s$
cancels out. For $N$ even there are other states cancelling the index.
Further examples of BPS states not contributing to the index can be found in
\deBoerRH\ eq. 5.10.


\subsec{Preliminary results on the basic $c=3$ theory $\CS$}

 Let us assume that the boson $\varphi$ in the $\CS$ theory has radius $R$.
Then the RR-sector supercharacter before GSO projection is
\eqn\youtooii{
Z_0 = \ \biggl\vert \SCh^{\CS,R}(\tau, \omega_+,\omega_-)
 \biggr\vert^2 Z_\Gamma}
where
\eqn\narain{
Z_{\Gamma}
=
\sum_{\Gamma^{1,1}} q^{\half p_L^2} \bar q^{\half p_R^2}
}
is a standard Siegel-Narain theta function for radius $R$.

The supercharacter is given by
\eqn\schr{
\eqalign{
&
\SCh^{\CS,R}(\tau, \omega_+,\omega_-)
 = q^{1/8} (z_+ + z_+^{-1} - z_- - z_-^{-1})\prod (1-q^n)^{-1} \cr
&
\prod_{n>0} (1- z_+ z_- q^n)(1- z_+^{-1} z_-^{-1} q^n)
(1- z_+^{-1} z_- q^n )(1- z_+  z_-^{-1} q^n )\cr}
}
The $q$ expansion of \schr\
defines a series of representations $R_n$ of $SU(2) \times SU(2)$ via:
\eqn\serrsp{
\SCh^{\CS,R} := q^{1/8} \sum_{n=0}^{\infty} q^n \chi_{R_n}(z_+,z_-)
}
where $R_n$ is a supercharacter of a (reducible)
 representation of $SU(2)\times SU(2)$.
It is useful to introduce the $SU(2)$ characters of
the irreducible representations of spin $\l$:
\eqn\sutwch{
\chi_\ell(z) := {z^{2\l +1} - z^{-2\l -1} 
\over z- z^{-1} } = z^{-2\l} + z^{-2\l + 2} +
\cdots + z^{+2\l}
}
In particular, we define
$u= \chi_{1/2}(z_+) $ and $ v=\chi_{1/2}(z_-)$.
Clearly, \schr\ vanishes at $z_+ = z_-$ and therefore
%
$
\chi_{R_n} = (u-v ) p_n(u,v) $
where $p_n(u,v)$ is a polynomial in $u,v$.

In our expansion   of $I_1$ in terms of $\cag$ characters
we will need the following crucial property:
 $\chi_{R_n}(z_+,z_-)$  has a first order zero iff
$n$ is a triangular number, that is, iff $n=\half m (m+1)$ for
some integer $m$. Moreover, for $n=\half m (m+1)$ we have
\eqn\exprssp{
p_{\half m (m+1)} (z_+,z_-) =(-1)^{m} \biggl(
 \chi_{m/2}(z_+) \chi_{m/2}(z_-) - 
\chi_{m/2-1/2}(z_+) \chi_{m/2-1/2}(z_-)\biggr)   + (u-v) q_m(u,v)
}
where $q_m(u,v)$ is a symmetric polynomial in $u,v$.

To prove this note that by applying \soperat\ to $\SCh^{\CS,R}$,
using the infinite product and then the infinite sum
representation of $\vartheta_1$ we find:
\eqn\thindx{
\eqalign{
-z_+ {d\over dz_-} \biggl\vert_{z_- =  z_+=z}
\SCh^{\CS,R} & =    
q^{1/8}  \sum_{m=0}^{\infty} 
q^{  \half m(m+1) } (-1)^m (z^{2m+1}  - z^{-2m-1})
  \cr}
}
Now note that
\eqn\chgvs{
-z_+ {d\over dz_-}
\biggl\vert_{z_- = z_+ = z}
\big[(u-v) p_n(u,v) \big] 
= (z-z^{-1}) p_n(u,u)
}
It thus follows from \thindx\ that $p_n(u,u)=0$ unless $n= \half m (m+1)$ is
a triangular number. Moreover, in the case that $n$ is triangular
\thindx\ and \chgvs\ implies
\eqn\trigns{
p_n(u,u) = (-1)^m {(z^{2m+1}  - z^{-2m-1}) \over z- z^{-1} }
}
Now note that putting $z_+ = z_-=z$ in \exprssp\
and using the Clebsch-Gordon decomposition we get $(-1)^m \chi_m(z)$, in
agreement with \trigns.

We conclude this subsection with an important remark.
The argument given above for   \exprssp\ only establishes
the existence of $q_m(u,v)$ as
a {\it virtual character} of $SU(2) \times SU(2)$.
Indeed, define
\eqn\defshi{
\varphi(j_1,j_2)
= \chi_{j_1}(z_+)\chi_{j_2}(z_-) - \chi_{j_1-1/2}(z_+)\chi_{j_2-1/2}(z_-).}
Observe that
\eqn\ambigu{
\varphi(j_1,j_2) = \varphi(j_1-1/2, j_2 + 1/2)  
+ (u-v) \chi_{j_1-1/2}(z_+) \chi_{j_2(z_-)}
}
This identity indeed establishes the leading term in the 
$q$-expansion of \mssmv.
Therefore, in general $ \varphi(j_1,j_2) 
= \varphi(j_1-j_3, j_2 + j_3) ~\mod~ (u-v)$.
The property that distinguishes the choice in \exprssp\ is that this is the
unique choice so that $\varphi(j_1,j_2)$ has a positive coefficient and
$q_m(u,v)$ is a {\it positive integer} combination of supercharacters.
This follows from an infinite product expansion for the character.
Note also that this is the unique choice which is symmetric
under $u\leftrightarrow v$.


\subsec{Computation of the index $I_2(\CC)$}

In order to compute the indices we need the following basic technical
result. The term   associated with $a,b,d$ in the sum
\teenn\ for $T_N$ is computed using
\eqn\contribinx{
\eqalign{
 - z_+ {d\over dz_-}\Bigl\vert_{z_- = z_+} 
\SCh^{\CS,R} & \Bigl({a \tau + b \over d},
a\omega_+ ,a\omega_-\Bigr)= \cr
& = a (z_+^a - z_+^{-a})  \qt^{1/8}
\prod_{n>0}  (1-\qt^n)  (1- z_+^{2a}   \qt^n)(1- z_{+}^{-2a} \qt^{n})  \cr
& = - i a \vartheta_1\Bigl(2a\omega, {a \tau + b\over d} \Bigr)
= a \Theta_{1,2}^-\Bigl(a\omega, {a \tau + b\over d} \Bigr) \cr
& =  a \sum_{n_0=0}^{d-1}  e^{4  \pi i {b\over d}
(n_0+{1\over 4})^2} \Theta_{a(4n_0+1), 2N }^- (\omega, \tau) \cr}
}
where  $N = ad$, and $z=z_+ =e^{2\pi i \omega}$, 
$\qt = \exp[2\pi i {a \tau + b\over d}]$
and we used the identities (D.7) and (D.8).
Applying \contribinx\ to both the left-movers and  the right-movers we get
\eqn\itwoindx{
\eqalign{
I_2({\rm Sym}^N(\CS)) &  = \sum_{ad=N} \sum_{n_0,m_0=0}^{d-1}   a
 \Theta_{a(4n_0+1),k}^-(\omega,\tau)
\overline{ \Theta_{a(4m_0+1),k}^-(\tl \omega,\tl \tau) }\cdot \cr
&\qquad \qquad
\cdot {1\over d} \sum_{b=0}^{d-1}e^{2\pi i {b \over d}(n_0-m_0)(2n_0+2m_0+1)}
Z_{\Gamma}({a\tau+b \over d}) \cr}
}
where $Z_\Gamma$ is defined in \narain.

The sum on $b$ becomes a projection operator, and depends on the charge sector
of the theory. Here we focus on the charge zero sector with $u=\tl u=0$.
In this case the index simplifies to
\eqn\splritwo{
I_2^0 = \sum_{ad=N} \sum_{n_0,m_0=0}^{d-1}   a \Pi^{(d)}(n_0,m_0)
 \Theta_{a(4n_0+1),k}^-(\omega,\tau) 
\overline{ \Theta_{a(4m_0+1),k}^-(\tl \omega,\tl \tau) }
}
where  we have introduced the projector
 $\Pi^{(d)}(n_0,m_0)$   onto
\eqn\projector{
(n_0-m_0) (2n_0 + 2m_0 + 1)  = 0 ~\mod~ d
}

Now we will assume $N$ prime. In this case there are only two contributions:
$a=N,d=1$ and $a=1,d=N$ in the sum over coverings. 
The ``short string'' contribution
 $a=N,d=1$
contributes $N \vert \Theta^-_{N,k}\vert^2$. The ``long string''
contribution $a=1,d=N$   is:
\eqn\aonden{
\sum_{n=0}^{N-1} \vert \Theta^-_{4n+1,k}\vert^2 +
\sum_{n=0, \mu\not=N}^{N-1} 
\Theta^-_{4n+1,k} \overline{ \Theta^-_{2N- (4n+1),k} }
}
Adding the contributions and using  
$\Theta^-_{\mu,k}= -\Theta^-_{2k-\mu,k}$ we arrive at
\itoosym.


\subsec{Computation of $I_1$ }

 Applying \contribinx\ only on the left-movers produces the   left-index:
\eqn\ioneindx{
\eqalign{
&
I_1({\rm Sym}^N(\CS)) = \sum_{ad=N} \sum_{n_0=0}^{d-1}
\Theta^-_{a(4n_0+1),k}(\omega,\tau)   \cr
& {1\over d} \sum_{b=0}^{d-1}  e^{4  \pi i {b\over d} (n_0+{1\over 4})^2}
\Biggl[\overline{ \SCh^{\CS,R} } 
\sum q^{\half p_L^2} \bar q^{\half p_R^2} \Biggr]
\Bigl({a \tau + b \over d} ; {a \bar \tau + b \over d} ,
a\tl\omega_+  , a\tl \omega_- \Bigr) \cr}
}

We can now extract the contribution $\Xi_\mu$ defined in \ione. Again
we will restrict attention to the charge zero sector and we will assume
$N$ is prime.

\subsubsec{\it Evaluation of the $a=N,d=1$ term}

We take the derivative in the index on the left.
Therefore we have to expand
\eqn\othrtrm{
\SCh^{\CS,R}(N\tau, N\omega_+, N \omega_-)
}
in terms of $\CA_{\gamma}$ characters for $k=2N$.
Note that   applying \contribinx\ to \othrtrm\ leads to
\eqn\stepone{
N \Theta_{N,k}^-(\omega,\tau)
}
Thus, we must find at least $N$ representations with
$\l^+ + \l^- = (N+1)/2$.

The $\cag$ highest weight states turn out to contribute only to the leading
term in the $q$-expansion of \othrtrm.
The spins of the BPS representations
can then be extracted by examining the $SU(2)\times SU(2)$
character in
this term and matching to the leading term in the $q$-expansion of the
characters of BPS representations of $\cag$. These leading terms are
summarized in appendix A.

The leading term in the $q$ expansion of \othrtrm\ is
\eqn\leding{
q^{N/8} \biggl( (\chi_{N/2}(z_+) - \chi_{(N-2)/2}(z_+) )
- (\chi_{N/2}(z_-) - \chi_{(N-2)/2}(z_-) \biggr)+ \cdots
}
In order to match this to the leading terms in the characters of $\cag$
 we need to factor out $u-v$ from the character in \leding.
This is  conveniently done using the relation of $SU(2)$ characters
to Chebyshev polynomials $U_n$.
This leads to   the key identity:
\eqn\keysint{
\chi_{1/2}(z_+^N) - \chi_{1/2}(z_-^N) = (u-v) \sum_{a + b = N-1}
\biggl[ \chi_{a/2}(z_+) \chi_{b/2}(z_-) - \chi_{(a-1)/2}(z_+)
\chi_{(b-1)/2}(z_-) \biggr]
}
where the sum is on integers $a,b\geq 0$, and we are using the identity
$\chi_{-1/2}(z)=0$.

After a little algebra one arrives at
\eqn\xpsdr{
\eqalign{
&
\Xi^{ss}_{N} = \SCh^{\CS,R}\Bigl(N\tau, z_+^N, z_-^N\Bigr) =
\sum_{\l^-=1/2}^{N/2} (-1)^{2\l^--1} \SCh({N+1\over 2}- \l^-, \l^-) +\cdots
\cr
}
}
The higher terms all involve massive characters with signs.

\example{$N=5$, $a=5$, $d=1$}
\eqn\nsgexmpl{\eqalign{
\Xi^{sg}_{\mu=5} & = \SCh(5/2,1/2) + \SCh(1/2,5/2) + \SCh(3/2,3/2) - \cr
& - \SCh(2,1) - \SCh(1,2) + \cr
& + \SChm (13/8, 5/2, 2/2) + \SChm (13/8, 2/2, 5/2) + \cr
& + \SChm (21/8, 5/2, 2/2) + \SChm (21/8, 2/2, 5/2) - \cr
& - 2 \SChm (21/8, 4/2, 3/2) - 2 \SChm (21/8, 3/2, 4/2) + \cr
& + \SChm (29/8, 5/2, 2/2) + \SChm (29/8, 2/2, 5/2) + \cr
& + 3 \SChm (29/8, 4/2, 3/2) + 3 \SChm (29/8, 3/2, 4/2) - \cr
& - 3 \SChm (37/8, 5/2, 2/2) - 3 \SChm (37/8, 2/2, 5/2) - \cr
& - 2 \SChm (37/8, 4/2, 3/2) - 2 \SChm (37/8, 3/2, 4/2) - \cr
& - 2 \SChm (45/8, 5/2, 2/2) - 2 \SChm (45/8, 2/2, 5/2) + \cr
& + \SChm (45/8, 4/2, 3/2) + \SChm (45/8, 3/2, 4/2) + \cr
& + \ldots
}}
Here, we introduced a shorthand notation,
$\SChm(h,\l^+,\l^-) := \SCh_{{\rm massive}} (h+{c \over 24},\l^+ ,\l^-)$.
Notice, that the massless supercharacters in the first two lines
agree with the decomposition \xpsdr.
However, the multiplicities of massive representations
do not have a definite sign.
\endexample

Comparing \xpsdr\ with the expansion \decopms\ shows that the
degeneracies $n^{ss}(\l^+,\l^-; \tl \l^+,\tl \l^-)$ vanish
unless $\mu=\tl \mu=N$, consistent with the index $I_2$.
Moreover, for $\tl \mu=N$ we have
\eqn\sumrule{
\sum_{\l^+ + \l^- = (N+1)/2} (-1)^{2\l^- + 1}
n^{ss}(\l^+,\l^-; \tl \l^+,\tl \l^-) = (-1)^{2\tl \l^- + 1}
}
for each representation with $\tl \l^+
= (N+1)/2-\tl \l^-$, $\half \leq \tl \l^- \leq N/2$.
The simplest spectrum consistent with the index is
\eqn\msslsr{
\bigoplus_{ } (\l^+,\l^-)_S \otimes \overline{ (\l^+,\l^-)_S }
}
where the sum is over all  spins, integer and half-integer with
\eqn\spinspr{
\l^++ \l^- = (N+1)/2, \qquad \l^\pm \geq 1/2
}
Again, we caution that \msslsr\ is only the simplest spectrum
consistent with the index $I_1$.


\subsubsec{\it Evaluation of the  $a=1, d=N$ term}

Now the ``long string''  contribution to $\Xi_\mu$ is
\eqn\leftyoutsp{
\Xi^{ls}_{\mu} := {1\over N} \sum_{b=0}^{N-1}
e^{-4\pi i {b\over N}  \bigl(n_0 + { 1\over  4}\bigr)^2}
   \SCh^{\CS,R }\Bigl({\tau+b \over N},  
\tilde \omega_+,  \tilde \omega_-\Bigr)
}
Here $\mu = 4n_0+1$, $n_0=0,\dots, N-1$. While $\Xi_\mu$ was defined in
\decopms\ only for $1\leq \mu \leq k-1$ we extend the range of
definition of $\mu$  in the same was as for $\Theta_{\mu,k}^-$.
Thus, $\Xi_{\mu + 4 N} = \Xi_{\mu}$ and $\Xi_{4N-\mu} = - \Xi_\mu$.
%
%

Now, we would like to decompose \leftyoutsp\ in terms of characters
of $\cag$ at level $(k^+,k^-)=(N,N)$ and in particular we would like to
isolate the contribution of the massless characters.
To this end we use the $q$-expansion \serrsp.
The projection from the sum over $b$ tells us that only the terms with
\eqn\projfs{
n=2n_0^2 + n_0 ~\mod~ N
}
survive in the sum \leftyoutsp\ and hence
\eqn\ximubh{
\Xi^{ls}_{\mu} = 
\sum_{n\geq 0, n=n_0(2n_0+1)~\mod~ N} e^{2\pi i {\tau\over N}(n+1/8) }
\chi_{\sst R_n}(z_+,z_-)
}
(We have used the definition \serrsp.)
Now, to extract the BPS representations we use the leading $q$ expansion of
BPS characters given in appendix \A. Accordingly, we look for $n$ such that
$\chi_{\sst R_n}$ has a first order zero at $z_+ = z_-$.
As discussed in sec. 4.3 this  happens
iff $n$ is a triangular number $n=\half m (m+1)$, in which case,
by \exprssp\
the $SU(2)\times SU(2)$ character is that of the leading
term in the $q$-expansion of the character of the
short $(\half m+\half, \half m+\half)$
representation of $\cag$.
Thus, a necessary condition for the $n^{th}$ term in \ximubh\ to correspond
to the leading term in the expansion of a BPS character is that
$n=\half m (m+1)$, in which case the spins must be
$(\half m+\half, \half m+\half)$. Note that such quantum numbers
indeed saturate the BPS bound:
\eqn\saturate{
k(L_0-c/24) = 2N {1\over N}(n+{1\over 8}) = (m+1/2)^2 = (\l^+ + \l^- - 1/2)^2
}

While there are infinitely many $n$ which satisfy the above criteria only
finitely many correspond to $\cag$ highest weight states since
 by the unitarity constraints we know that $0 \leq m\leq (N-1)$.
Moreover, by the projection \projfs\ we have
\eqn\pronjs{
(m-2n_0)(m+2n_0+1) = 0 ~\mod~ N
}
We now solve these constraints:
For $0\leq n_0 \leq (N-1)/2$ the only solutions are
$m=2n_0$ and $m=(N-1)-2n_0$. For $(N+1)/2\leq n_0 \leq (N-1)$
the only solutions are $m=2n_0-N$ and $m=(N-1)-(2n_0-N)$.
This shows which BPS representations {\it might} occur. On the other
hand, the index $I_2$ shows that these representations, at least, must
occur.

Thus, we conclude that for 
$0\leq n_0 \leq (N-1)/2$, and $n_0 \not= (N-1)/4$ we have
\eqn\expbhxi{
\Xi^{ls}_{\mu} = \SCh(n_0+1/2, n_0+1/2) + \SCh\Bigl({N-1\over 2}
-n_0 +1/2,{N-1\over 2} -n_0 +1/2\Bigr) + \cdots
}
Note that this corresponds to the range  
$1\leq \mu \leq 2N-1$, $\mu=1~\mod~ 4$,
and $\mu\not= N$, and can be expressed as:
\eqn\expbhxi{
\Xi^{ls}_{\mu} = \SCh\Bigl({\mu+1\over 4},{\mu+1\over 4} \Bigr) +
\SCh\Bigl({N+1\over 2} -{\mu+1\over 4},{N+1\over 2} -{\mu+1\over 4}\Bigr) 
+ \cdots
}

For $(N+1)/2\leq n_0 \leq (N-1)$, and $n_0 \not= (3N-1)/4$ we have
\eqn\expbhxii{
\Xi^{ls}_{\mu} = \SCh\Bigl(n_0-{N\over 2} +1/2, n_0-{N\over 2} +1/2\Bigr)
+ \SCh(N-n_0,N-n_0) + \cdots
}
Note this corresponds to the range $2N+3\leq \mu \leq 4N-3$, $\mu=1~\mod~ 4$,
$\mu\not= 3N$. It is more convenient to work with $\mu$ in the original
range $1\leq \mu\leq 2N-1$, so we use $\Xi_{\mu}= - \Xi_{4N-\mu}$ to say
\eqn\expbhxii{
\Xi^{ls}_{\mu} = -\SCh\Bigl({\mu-1\over 4},{\mu-1\over 4} \Bigr) -
\SCh\Bigl({N+1\over 2} -{\mu-1\over 4},{N+1\over 2} -{\mu-1\over 4}\Bigr) 
+ \cdots
}
where now  $3 \leq \mu \leq (2N-3)$, $\mu= 3 ~\mod~ 4$, and $\mu \not=N$.
Finally, if $\mu=N$ we have
\eqn\muennbh{
\Xi^{ls}_{\mu=N} = \epsilon \SCh\Bigl({N+1\over 4},{N+1\over 4} \Bigr) + \cdots
}
where $\epsilon = (-1)^{\half N (N-1)}$.

It thus follows that there are nonzero degeneracies for all representations
$(\l,\l)$, $1/2 \leq \l \leq N/2$.
%
%
The simplest spectrum consistent with the index is
\eqn\spimop{
\bigoplus_{\l=1/2}^{(N-1)/4} \biggl\vert (\l,\l)
+ \Bigl({N+1\over 2}-\l, {N+1\over 2}-\l\Bigr) \biggr\vert^2
\oplus \biggl\vert  \Bigl({N+1\over 4} , {N+1\over 4} \Bigr) \biggr\vert^2
}
where all spins, integer and half-integer appear.

Again, note that the  arguments above only establish that the terms
$+\cdots$ above are linear combinations of massive supercharacters with signs.
However, because of \mssmv\ this leaves ambiguous the short representations
which appear. This ambiguity may be removed as follows. Note, that from
the index $I_2$ the spectrum of massless representations is diagonal up to
reflection of spins.
This property can be used to constrain the multiplicities of
both massless and massive supercharacters that appear in
the decomposition of $\Xi^{ls}_{\mu}$.
Indeed, from \decopms\ we find that the multiplicities of
massless supercharacters, $\SCh(\tl \l^+,\tl \l^-)$, are given by
$(-1)^{{\mu + 3 \over 2}} n(\l, \l ;\tl \l , \tl \l)~\delta_{\mu,4\l-1}$,
whereas the multiplicities of massive supercharacters,
$\SCh(\tl \rhom)$, are given by
$(-1)^{{\mu + 3 \over 2}} N(\l,\l; \tl \rhom)$.
Since the degeneracies $N(\rho,\tl \rho)$ in \fulldec\ are {\it positive},
we conclude that \expbhxi\ should be expanded in {\it positive integral}
combinations of supercharacters,  while \expbhxii\ should be expanded
in negative integral combinations of supercharacters.
We verified this property (by computer) in a number
of examples.\foot{The computer code is available upon request.}

\example{$N=5$, $a=1$, $d=5$}
\eqn\nlsexmpl{\eqalign{
\Xi^{ls}_{\mu=9} & = \SCh(1/2,1/2) + \SCh(3/2,3/2) + \cr
& +  3 \SChm (41/40, 3/2, 4/2) + 3 \SChm (41/40, 4/2, 3/2) + \cr
& + 6 \SChm (41/40, 3/2, 2/2) + 6 \SChm (41/40, 2/2, 3/2) + \cr
& + 6 \SChm (81/40, 5/2, 4/2) + 6 \SChm (81/40, 4/2, 5/2) + \cr
& + 10 \SChm (81/40, 5/2, 2/2) + 10 \SChm (81/40, 2/2, 5/2) + \cr
& + 42 \SChm (81/40, 4/2, 3/2) + 42 \SChm (81/40, 3/2, 4/2) + \cr
& + 61 \SChm (81/40, 3/2, 2/2) + 61 \SChm (81/40, 2/2, 3/2) + \cr
& + 61 \SChm (121/40, 5/2, 4/2) + 61 \SChm (121/40, 4/2, 5/2) + \cr
& + 75 \SChm (121/40, 5/2, 2/2) + 75 \SChm (121/40, 2/2, 5/2) + \cr
& + 290 \SChm (121/40, 4/2, 3/2) + 290 \SChm (121/40, 3/2, 4/2) + \cr
& + 348 \SChm (121/40, 3/2, 2/2) + 348 \SChm (121/40, 2/2, 3/2) + \cr
& + \ldots
}}
Notice, that massless supercharacters in the first line agree with
the expansion \expbhxi, and that the multiplicities of massive
representations are all positive and form a regular pattern.
\endexample

We should stress, however, that this method allows to completely
remove the ambiguity only in the $a=1$, $d=N$ term, where $\Xi_{\mu}$
decomposes into massless and massive supercharacters of a definite sign.


\subsec{Charged excitations}

Let us now indicate how the results generalize to the spectrum of
charged excitations in the $\syms$ theory.

If the radius of the boson in the $\CS$ theory is finite,
$\varphi \sim \varphi + 2\pi R$, then there will be
a Narain lattice of charges. A state with momentum
$n$ and winding $w$ contributes:
\eqn\naraii{
q^{{\alpha'\over 4} (n/R + wR/\alpha')^2}
\bar q^{{\alpha'\over 4} (n/R - wR/\alpha')^2}
}
In units with $\alpha'=2$
the spectrum of $(p_L,p_R)$  is
\eqn\plpr{
(p_L,p_R)= (n/R + wR/2, n/R - wR/2)
}
with $n,w\in \IZ$. Let us call this lattice $\Gamma(R) \subset \IR^2$.

  Given that the charges $(u_L; u_R)$ of the
$\CS$ theory form a Narain lattice $\Gamma(R)$ we would
like to describe the lattice of charges
\eqn\chargelattice{
\Gamma_{\syms} := \Bigl\{ (u_L; u_R) \vert \exists \psi \in \CH(\syms)
~{\rm s.t.}~  U\psi = u_L \psi, \tilde U \psi= u_R \psi \Bigr\}
}
of the $\syms$ theory, and how those charges are correlated with uncharged
states. It is easy to see that
the charges $(u_L; u_R)$ always lie in 
the same Narain lattice $\Gamma(R)$ as the
parent theory $\CS$, i.e. $\Gamma_{\syms}\subset \Gamma(R)$.   
We now discuss the
charged states in more detail.

The conventionally normalized $U(1)$ current in the $\cag$ algebra is
\eqn\prop{
U(z) U(w) \sim - {k\over 2} {1\over (z-w)^2} + \cdots
}
In the $\CS$ theory there is a scalar field $\varphi$ and
$U(z) = \p \varphi$ with $k=2$. In the $\syms$ theory we can form
the field 
$\varphi_{\rm \sst Diag}: = \varphi_1 + \varphi_2 + \cdots + \varphi_N$.
Notice that $\varphi$ and $\varphi_{\rm \sst Diag}$ 
have the {\it same} periodicity,
$$
\varphi_{\rm \sst Diag} \sim \varphi_{\rm \sst Diag} + 2\pi R
$$
and that
$$
U := \p \varphi_{\rm \sst Diag}
$$
is the conventionally normalized $U(1)$ current in the $\cag$ algebra of
$\syms$  because
$k=2N$. It follows that in the untwisted sector the spectrum of
charges  is $N \Gamma(R)$. Moreover, let us note that we may bosonize
the current $U$ by introducing a conventionally normalized
effective boson $\Phi_{\rm eff}$ by $U =  \sqrt{N} \p \Phi_{\rm eff}$.
Clearly $\Phi_{\rm eff} = {1\over \sqrt{N}} \varphi_{\rm \sst Diag}$ 
and hence it has
radius $R/\sqrt{N}$. Thus, the radius of $\Phi_{\rm eff}$ is
determined by the radius of $\varphi$.

All operators in $\syms$  can be written in the form
\eqn\chgdop{
\Phi = \Phi_0 e^{i {u_L\over \sqrt{N}} 
\Phi_{\rm eff}} e^{i {u_R\over \sqrt{N}} \tilde \Phi_{\rm eff}}
}
where $\Phi_0$ is neutral under $U,\tilde U$ and
\eqn\chgedgap{
\eqalign{
h & = h_0 + {u_L^2\over 2N} \cr
\tilde h & = \tilde h_0 + {u_R^2\over 2N} \cr}
}
However, which operators $\Phi_0$ appear 
is correlated with $(u_L;u_R)$ in a complicated
way. Moreover, $\Phi_{\rm eff}\to \Phi_{\rm eff} 
+ 2\pi  R/\sqrt{N}$ is not a global symmetry of
the theory, but must be accompanied by a discrete 
transformation on the fields $\Phi_0$.

Equation \chgedgap\ is true for any 
operators in $\syms$, BPS or not. However, from the
BPS mass formula, we can say that \chgedgap\ implies that 
$\Phi$ is BPS iff $\Phi_0$ is BPS.

We can gain some further insight about which charged BPS states do occur
by closer examination of  \itwoindx. In the $a=N,d=1$ term
we find a multiplicative factor $Z_{\Gamma}(N\tau)$. These contribute
terms
$$
q^{{N\over 2} p_L^2} \bar q^{{N\over 2} p_R^2} =
q^{{1\over 2N} (Np_L)^2} \bar q^{{1\over 2N} (Np_R)^2}
$$
In the second equality we have written the weights in a form such that, from
the BPS formula $h = h_0 + u^2/k$ we can read off the properly normalized
charges. This is in accord with the claim 
that the untwisted sector charge lattice is
$N \Gamma(R)$.
%
%
When $d>1$ the sum over $b$ induced a nontrivial projection, which
in general correlates states with different $n_0,m_0$ with the charged
excitation in a complicated way. 
The projection condition  \projector\ is modified to
\eqn\projectorp{
(n_0-m_0) (2n_0 + 2m_0 + 1)  + nw =0 ~\mod~ d
}
where $n,w$ are related to $(p_L;p_R)$ by \plpr.
 The spectrum of $(bps,bps)$ states in the $N=ad$ sector is of the form:
\eqn\specfrm{
\eqalign{
h & = {a^2(4n_0+1)^2\over 8N} + N_L + {1\over 2N} (ap_L)^2 \cr
\tilde h & = {a^2(4m_0+1)^2\over 8N} + N_R + {1\over 2N} (ap_R)^2 \cr}
}
where $N_L, N_R \in \IZ_+$, and $h-\tilde h \in \IZ$ is equivalent to
\projectorp. For $d>1$ there is a nontrivial correlation,
given by \projectorp\ of which $\Phi_0$ can occur with which Narain vectors.

All charged $(bps,bps)$ states have gaps 
relative to uncharged states of the form
\eqn\gaps{
\eqalign{
h & = h_0 + {p_L^2\over 2N} \cr
\tilde h & = \tilde h_0 + {p_R^2\over 2N} \cr}
}
where $(p_L;p_R)$ is {\it some} vector in $\Gamma(R)$. This should be
carefully distinguished from the statement that for {\it any}
$(p_L;p_R)$ and $\Phi_0$ there is a corresponding $\Phi$ with
gap  \gaps. Indeed, we want to stress that no local quantization of a
single Gaussian field can give a spectrum of conformal weights of the form
\eqn\fakenarain{
(h,\tilde h) = (
{p_L^2\over 2N}  , {p_R^2\over 2N} )
}
The spectrum  \fakenarain\ for $(p_L; p_R) \in \Gamma(R)$
is not equivalent to the spectrum obtained from any Narain
lattice at any  other radius $R'$.

Finally, let us note that the
  contribution of the charged states to the index can be
systematically discussed by using the extended   definition of $Z$ in
\chargdfnze. In terms of the $\CS$ theory we replace
\eqn\repsl{
Z_\Gamma \to Z_\Gamma(\tau,\chi) := \sum_{p\in \Gamma(R)}
 e^{i \pi \tau p_L^2 - i \pi \bar \tau p_R^2
+ 2\pi i \chi\cdot p }
}
Note that \repsl\ is a  Siegel-Narain
theta function for the lattice $II^{1,1}$ embedded in $\IR^{1,1}$ according
to \plpr.

Now, in evaluating the indices $I_1$ and $I_2$, $Z_\Gamma$ enters
through the combination
\eqn\prjsn{
{1\over d} \sum_{b=0}^{d-1} 
e^{2\pi i \nu {b\over d}} Z_{\Gamma}\Bigl({a \tau + b\over d},a\chi\Bigr)
}
for various values of $\nu$.
We  claim that \prjsn\
can be expressed in terms of Siegel-Narain theta functions of higher level.
We can write
\prjsn\ as
\eqn\sumfirn{
\sum_{m,n} \delta_{mn = - \nu ~\mod~ d} \cdot
\exp\biggl[ i \pi \tau{ 1\over N} (a p_L)^2 - 
i \pi \bar \tau {1\over N}(a p_R)^2 + 2\pi i a \chi\cdot p \biggr]
}
where  $(p_L;p_R) = m e + nf $ and $e=  (1/R; 1/R)$,
$f= {1\over 2}(R; -R)$. We can view this as a  sum over lattice vectors:
\eqn\lattvectrs{
v = {a\over \sqrt{N}}p = {a\over \sqrt{N}} m e + {a\over \sqrt{N}} n f
}
where $mn=-\nu ~\mod~ d$. Let $(m_0^i, n_0^i)$
run over the  the finite set of distinct solutions of 
this congruence, modulo $d$. Then
the general solution is $m=m_0^i + \ell_1 d , n= n_0^i + \ell_2 d $
where $\ell_1,\ell_2\in \IZ$. Thus, we define
\eqn\lattice{
\Lambda = \sqrt{N} e \IZ + \sqrt{N} f \IZ = \sqrt{N}\Gamma(R)
}
and note that $v = \beta + \lambda$, $\lambda \in \Lambda$, and
$\beta \in a\Lambda^*$ satisfies
\eqn\betaconstr{
{N\over 2} \beta^2 = - a^2 \nu ~\mod~ (a N)
}
We can thus write \prjsn\ in terms of Siegel-Narain theta functions
for the lattice $\Lambda$. Including the exponential prefactor in
\chargdfnze\ we have
\eqn\prsjnres{
e^{{N\pi \over 2 \tau_2} (\chi_L^2+\chi_R^2)}
{1\over d} \sum_{b=0}^{d-1} e^{2\pi i \nu {b\over d}} 
Z_{\Gamma}\Bigl({a \tau + b\over d}\Bigr)
= \sum_{\beta \in a \Lambda^*/\Lambda} \Pi_a(\beta,\nu)  
\Theta_\Lambda(\tau,0,\beta;P;\sqrt{N} \chi )
}
where $\Pi_a(\beta,\nu)$ projects onto solutions of \betaconstr\ and
$\Theta_\Lambda(\tau,\alpha ,\beta;P;\xi)$ 
is a Siegel-Narain theta function for
$\Lambda$.
It follows that we can summarize the $\chi$-dependence in  $I_2$ according to
\eqn\itwochrgp{
I_2 = \sum_{\beta \in \Lambda^*/\Lambda} I_2^\beta  
\Theta_\Lambda(\tau,0,\beta;P;\sqrt{N} \chi )
}
where  $I_2^\beta$ are $\chi$-independent and
can be written  in terms of level $k$ theta functions of $\omega,\tl \omega$,
in expressions generalizing $I_2^0$ in \itoosym. (This was our motivation
for introducing \itwochrg\ above.)
Using \prsjnres\ one finds that similar
  results hold for $I_1$. These expressions are of interest
in the AdS/CFT context  because in
the holographic dual theory, the $U(1)\times U(1)$ Chern-Simons sector at level $Q_1$
is naturally expressed as a linear combination of level $Q_1$ Siegel-Narain
theta functions \refs{\GukovYM,\GukovID}.


\subsec{Moduli of the $\syms$ theory}

In \GukovYM\   criteria for operators to induce moduli spaces of
theories with $\cag$ symmetry are investigated.
We need a short multiplet which
contains a state with $h=\tilde h=1 $ and which is a singlet under
the $R$-symmetry $SU(2)^4$. This only occurs as a state in the short NS
representations $(0,0)$ and $(\half,\half)$ of $\cag$.

Examination of the index yields two such moduli. First,
 $\sum_i \p \varphi(i) \pb \varphi(i) $  is the modulus
corresponding to deformations of the radius $R$ of the $\CS$ theory.
This is the state in the $(\half,\half)$ representation, predicted by
the index. Second,  the vacuum representation  
$(0,0)$ is always present.  This
leads to a universal modulus $U_{-1} \tilde U_{-1} \vert 0 \rangle $,
that is $\sim \p \Phi_{\rm eff} \pb \Phi_{\rm eff}$.  Perturbation by
this operator formally
deforms the radius of $\Phi_{\rm eff}$, but we have seen that at the
orbifold point the radius of $\Phi_{\rm eff}$ is
not independent of $R$!  In fact, the deformation by this operator
does not preserve the structure of a symmetric product orbifold
(as is easily confirmed by studying first order conformal perturbation
theory).

 It is important to recognize that     some BPS
states exist at the symmetric product point, but do not contribute
to the index. Among these there is a potential modulus in the
$n=3$ term in \twistste.
Nevertheless,  we can invoke the general theorem proved in \GukovYM\ to see
that such a state is a true modulus.%
\foot{ It might be useful to note that the marginal operator can
be written simply as follows. Consider the state
$\vert \epsilon_{\mu\nu\lambda\rho}  
\psi^\mu \psi^\nu \psi^\lambda \psi^\rho \vert^2
\vert 0 \rangle_{NS}$ in the $3$-long string sector $\CH^{(3)}$. This state
corresponds to the marginal deformation. }
This state therefore survives deformations as a BPS state, even though
it does not contribute to the index!

Finally, there   are potential extra moduli arising from charged BPS states
when the radius $R$ takes special values.     For example if $p_L^2= p_R^2$
we get extra contributions to the spectrum.
 At special radii these can give moduli, e.g.
if $u^2=k$ then the massless NS rep  $(h=u^2/k =1 , \l^+=0,\l^-=0)_S$
can represent a modulus.

Thus, we conclude that the $\syms$ theory generically 
has three moduli, only two of
which are detected by the index.


\subsec{Comments on $N$ not prime}

%

\subsubsec{\it Computation of $I_2$}

As explained above,
in the zero charge sector of the Sym$^N(\CS)$ theory
the index $I_2$ is given by the sum over factorizations of $N$.
Hence, if $N$ is not prime, there are additional contributions
to the index. Our main motivation for doing this computation
was the AdS/CFT application discussed in \GukovYM. In particular,
we were interested in comparing with low lying supergravity
states in the case where $N$ is not prime.

To begin, let us consider a contribution associated
with the factorization $N=ad$, where both $a$ and $d$ are prime.
It follows from \projector\ that there can be two possibilities:
$1)$ $(n_0 - m_0)$ is divisible by $d$, or
$2)$ $(2n_0 + 2m_0 + 1)$ is divisible by $d$.
Using the fact that both $m_0$ and $n_0$ take values in the range
$0, \ldots, (d-1)$, it is easy to find the allowed values
of $\mu$ and $\tilde \mu$ in each case:
\eqn\solonetwo{\eqalign{
& 1) \quad \quad \mu = a(4n_0+1) \quad \quad \tilde \mu = a(4n_0+1)  \cr
& 2a) \quad \quad \mu = a(4n_0+1) \quad \quad \tilde \mu = a(2d - (4n_0+1)) \cr
& 2b) \quad \quad \mu = a(4n_0+1) \quad \quad \tilde \mu = a(6d - (4n_0+1))
}}
According to \itwoindx, these values of $(\mu , \tilde \mu)$
occur with multiplicity $a$ in the massless spectrum
of the Sym$^N (\CS)$ theory.
They determine multiplicities, 
$n (\ell^+ , \ell^- ; \tilde \ell^+ , \tilde \ell^-)$,
of massless representations in the zero charge sector via \degnes.
The resulting contribution to the index is similar to
the ``long string'' contribution \aonden:
\eqn\aondenp{
a \sum_{n=0}^{d-1} \vert \Theta^-_{a(4n+1),k}\vert^2
+ 2a \sum_{n=0, \mu\not=N}^{d-1} \Theta^-_{a(4n+1),k}
\overline{ \Theta^-_{2N- a(4n+1),k} }
}
where writing the last term we used
$\Theta^-_{\mu,k}= \Theta^-_{4N+\mu,k}$.
Notice, in particular, that the index $I_2$ requires states in
the NS spectrum with small values of $(\ell^+ - \ell^-)$,
which might possibly modify the supergravity spectrum
when $N$ is not prime. To see this, let's start with
a massless state\foot{Here, we consider only the left
sector. The right sector is completely similar.} $(\ell^+ , \ell^-)$
in the NS sector with small values of $\ell^+$ and $\ell^-$.
Under the spectral flow, this state transforms into
a state $(N/2 - \ell^+ , \ell^- + 1/2)$.
According to \degnes, it corresponds to
$$
\mu = \tilde \mu = N + 2 (\ell^- - \ell^+)
$$
Since $\ell^{\pm}$ are assumed to be small compared to $N$,
we have $\mu = \tilde \mu \simeq N$, which is allowed
according to \solonetwo. Therefore, the index $I_2$ alone
does not rule out a possibility that there are extra
supergravity states when $N$ is not prime.
In order to obtain more information, let us look at
the more refined index $I_1$.


\subsubsec{\it Computation of $I_1$}

Generalizing the results for $I_1$ takes more effort.
Let us start with the contribution from the factorization $N= ad$
with $\mu=\tl\mu$, i.e. the contributions with $n_0=m_0$ in \projector.
In this case, $\mu = a (4n_0+1)$, and we are looking at BPS
representations with $\tl \l^+ + \tl \l^-  = (\mu+1)/2$.
We expand $\Xi_\mu$ extracted from \ioneindx\ in a $q$ expansion,
and find, from the BPS bound $\tl L_0 - c/24 = \mu^2/(4k)$
corresponding to the $n^{th}$ term in the expansion
with $n= \half (2n_0)(2n_0+1)$.
In order to obtain the corresponding spectrum of massless representations,
we need to find the coefficient of the $(u-v)$ term
in the decomposition of $\chi_{R_n}(z_+^a, z_-^a)$
in terms of $SU(2) \times SU(2)$ characters.
After some algebra, we find
\eqn\beeta{
\eqalign{
\chi_{R_n}(z_+^a, z_-^a) = (u-v) \sum_{\alpha+\beta= a-1}
&
\biggl( \chi_{an_0 + \alpha/2 }(z_+) \chi_{an_0 + \beta/2}(z_-) \cr
&  -
\chi_{an_0 + \alpha/2 -1/2}(z_+) \chi_{an_0 + \beta/2-1/2 }(z_-)\biggr)
~\mod~ (u-v)^2\cr}
}
which leads to the following spectrum of massless representations:
\eqn\minspect{
\bigoplus_{n_0=0}^{d-1}\bigoplus_{\alpha+\beta=(a-1)}
 \Bigl\vert ( an_0 + (\alpha +1)/2 , a n_0  + (\beta +1)/2 )\Bigr\vert^2
}
As before, this spectrum is not unique; this is related to
the non-uniqueness of the coefficient of $(u-v)$ in eq. \beeta.
We chose the particular form of \beeta\ because it is manifestly
symmetric under exchange $z_+ \leftrightarrow z_-$,
and because the resulting spectrum \minspect\ is a natural
generalization of what we found for $a=N,d=1$ and $a=1,d=N$.

The states \minspect\ have conformal weights
\eqn\lognwt{
h= {N\over 8 } + {a^2 (4n_0+1)^2\over 8N} .
}
Upon
spectral flow to the NS sector these map to states $\vert (j_+,j_-)\vert^2$
with
\eqn\quanm{
\eqalign{
j^+ - j^- & = \half (N-\mu) \cr
h & = {1\over 4} (N+a - 2 \alpha -2)
+ {1\over 8N}(N- \mu)^2\qquad 0 \leq \alpha \leq a-1 \cr}
}
where $\mu = a ~\mod~ 4a$, i.e. $\mu = a, 5a, 9a,\dots, 4N-3a$.
Note we can write the conformal weight as
\eqn\notgns{
h = {(N-1)\over 4} + {(N-\mu)^2\over 8N} +  
{a - (\alpha+1)\over 2} - {a-1\over 4}
}
Comparison with   \lngwt\ shows that
 this is in a sense a composite of the short string and long string spectrum.
The short string spectrum is the contribution ${a - (\alpha+1)\over 2}$ for
$0 \leq \alpha\leq a-1$.  This adds support to the  picture that $V(p)$
adds shorts strings to a background of long strings created by $U(p)$.

There are other contributions in the $N=ad$ term, from
off-diagonal terms with $\mu\not=\tl \mu$.
Consider the $(a,d)$-contribution to $\Xi_{\mu}$:
\eqn\xis{\eqalign{
\Xi_{a(4n_0 + 1)} & \equiv
\sum_{{\ell^+ + \ell^- = (\mu + 1)/2 \atop \tl \l^+ , \tl \l^-}}
(-1)^{2\ell^- + 1}
n (\ell^{\pm}; \tilde \ell^{\pm}) {\rm ~SCh}_{(\tilde \ell^+, \tilde \ell^-)}
+ {\rm ~massive} = \cr
& = \sum_{n \in \CN} q^{{a \over d} (n + 1/8)} \chi_{R_n} (z_+^a , z_-^a)
}}
Here, $\CN$ is a set of triangular numbers which satisfy
the following relation:
\eqn\nnnrel{ n = n_0 (2n_0 + 1) \quad {\rm mod}~~d }
Writing $n = m(m+1)/2$, from \nnnrel\ we get
\eqn\mnrel{ {1 \over 2} (m-2n_0) (m + 2n_0 + 1) = 0 \quad {\rm mod}~~d }
As in the previous case, our goal here is to describe
the set $\CN$ explicitly. As in \solonetwo,
when $d$ is prime ($d \ne 2$) we find two groups of solutions
corresponding to the two factors in \mnrel:
\eqn\mnsol{\eqalign{
& 1) \quad \quad  m=2n_0 + sd \quad\quad s=-1,0,1, \ldots, (a-1)  \cr
& 2) \quad \quad  m=sd-2n_0 -1 \quad\quad s=1,2, \ldots, (a+1)
}}
where, depending on the value of $s$, $n_0$ runs over
\eqn\nrun{\eqalign{
& 1) \quad\quad \max \left( 0 , -{sd \over 2}\right)
\le n_0 \le \min \left( d-1 , {(a-s)d -1 \over 2}\right) \cr
& 2) \quad\quad \max \left( 0 , {(s-a)d \over 2}\right)
\le n_0 \le \min \left( d-1 , {sd -1 \over 2}\right)
}}
Notice, that except for the largest or smallest values of $s$,
$n_0$ takes values in its natural range $0 \le n_0 \le (d-1)$.

Therefore, we can label the elements of the set $\CN$
by paris $(n_0,s)$ in the range \mnsol\ -- \nrun.
Using the decomposition \beeta\ of $\chi_{R_n}(z_+^a , z_-^a)$
in terms of the $SU(2) \times SU(2)$ characters:
\eqn\chirndecomp{
\chi_{R_n}(z_+^a , z_-^a) \to
\bigoplus_{\a + \b = (a-1)}
\left({am + \a + 1 \over 2} ~,~ {am + \b + 1 \over 2} \right)
}
and substituting it into \xis, we find that
%
%
%
the simplest spectrum consistent with the index $I_1$ is,
{\it cf.} \minspect,
\eqn\specp{
\bigoplus_{{(n_0,s) \atop \a + \b = (a-1)}}
\left({2an_0 + \a + 1 \over 2} ~,~ {2an_0 + \b + 1 \over 2} ;
{a(2n_0 + sd) + \a + 1 \over 2} ~,~ {a(2n_0 + sd) + \b + 1 \over 2} \right)
}
$$
\bigoplus_{{(n_0,s) \atop \a + \b = (a-1)}}
\left( {2an_0 + \a + 1 \over 2} ~,~ {2an_0 + \b + 1 \over 2} ;
{a(sd-2n_0 -1) + \a + 1 \over 2} ~,~ {a(sd-2n_0 -1) + \b + 1 \over 2} \right)
$$
Note, that this spectrum does not contain states
with $\ell^+ \simeq N/2$ and $\ell^- \simeq 1$, which under spectral
flow might become a part of the supergravity spectrum.
The reason is simply that the difference $(\ell^+ - \ell^-)$
for the states \specp\ can not exceed $(\a + \b)/2 \simeq a/2 \le N/4$.

Finally, we point out that, when $d=2$, the analysis is very similar.
Namely, one finds that the spectrum of massless states can still be
described by \specp\ where only even values of $s$ appear.

\bigskip {\noindent {\it{Evaluation for $N=p^r$}}}

Now let us discuss the case when $a$ and $d$ are not relatively prime.
In particualar, we will consider the situation when $a$ and $d$ contain
several prime factors. The simplest example of this is when $N$
is of the form:
\eqn\npr{N = p^r}
Then, the index can be written as a sum over
factorizations of $N$, which in this case involve factors like
\eqn\adnpr{d = p^{\d} \quad , \quad a = p^{r-\d} }
for some $\d = 0, \ldots, r$.

As in the case of $a$ and $d$ prime, the left index $I_1$
has the form \ioneindx, where $a$ and $d$ are now given by \adnpr\ and
\eqn\munpr{
\mu = p^{r-\d} (4n_0 + 1)
}
Specifically, we can write \ioneindx\ as a sum over $\d$
\eqn\ionenpr{
I^0_1 = \sum_{\d = o}^{r}\sum_{n_0 = 0}^{p^{\d}-1} \Theta^-_{\mu,2 p^r}
\times {1 \over p^{\d}} \sum_{b=0}^{p^{\d}-1} 
e^{4\pi i {b \over d} (n_0 + 1/4)^2}
{\bar {\rm ~SCh}^{\CS,R}} \left( {a\tau + b \over d} , z_+^a , z_-^a \right)
}
where a contribution from each value of $\d$ looks like
\eqn\xiss{
\Xi_{\mu} =
\sum_{n \in \CN} q^{{a \over d} (n + 1/8)} \chi_{R_n} (z_+^a , z_-^a)
}
The sum here is over triangular numbers $n = m(m+1)/2$,
which satisfy the condition \mnrel:
\eqn\mnnpr{ {1 \over 2} (m-2n_0) (m + 2n_0 + 1) = 0 \quad {\rm mod}~~p^{\d} }
For the moment let us assume that $p \ne 2$. Then, equation \mnnpr\
implies that the product on the left-hand side is divisible by $p^{\d}$.
This can happen only if each factor is divisible by a certain power of $p$:
\eqn\ppdiv{\eqalign{
& (m-2n_0) \quad {\rm ~divisible~by~} \quad p^{\d_1} \cr
& (m+2n_0+1) \quad {\rm ~divisible~by~} \quad p^{\d_2}
}}
such that $\d_1 + \d_2 = \d$ and $\d_{1,2} \ge 0$.
There are two types of solutions we have to consider:
\eqn\nprcases{\eqalign{
& 1) \quad \quad \d_1 = 0 \quad {\rm ~or~} \quad \d_2 =0 \cr
& 2) \quad \quad \d_1 > 0 \quad , \quad \d_2 > 0
}}
The first case is essentially the same as the one discussed in the previous
subsection, when $a$ and $d$ are prime. In particualar, we find the same
spectrum \specp, where the values of $a$ and $d$ are given by \adnpr.

The second case in \nprcases\ describes extra states which appear when
$a$ and $d$ are not prime, and this is really what we are after.
In this case, we can write \ppdiv\ more explicitly as
%
\eqn\pppmn{\eqalign{
2 m & = -1 + s_1 p^{\d_1} + s_2 p^{\d_2} \cr
4 n_0 & = -1 - s_1 p^{\d_1} + s_2 p^{\d_2}
}}
In particular, these equations imply $p \vert (2m+1)$ and $p \vert (4n_0 + 1)$.

Notice, that for $p=2$, instead of \ppdiv\ we find that either $(m-2n_0)$
or $(m+2n_0 + 1)$ should be divisible by $2^{\d+1}$. There is no mixed
case like $2)$ in \nprcases\ because $(m-2n_0)$ and $(m+2n_0 + 1)$ can not
be both even.
Therefore, we conclude that, for $p=2$, there are no extra massless
states in the spectrum, except for those already describes in \specp.

Hence, without loss of generality, in what follows we consider $p \ne 2$,
which is the case when extra massless states {\it can} occur.
In order to identify these states, we need to find integer solutions to \pppmn.
First, for given $\d_1$ and $\d_2$,
let us consider the range of values of $s_1$ and $s_2$,
which are consistent with the bounds on $m$ and $n_0$:
$0 \le n_0 \le d-1$ and $0 \le m \le ad-1$. From \pppmn\ we find
%
\eqn\ssbound{\eqalign{
& -2p^{\d - \d_1} \le s_1 \le p^{r-\d_1} -1 \cr
& 1 \le s_2 \le 2 p^{\d-\d_2} + p^{r-\d_2} -1
}}
Therefore, we conclude that $s_1$ and $s_2$ in this range,
which solve \pppmn\ for integer values of $m$ and $n_0$,
lead to extra massless states of the form
\eqn\extrastates{
\bigoplus_{{(n_0,\d_{1,2}, s_{1,2}) \atop \a + \b = (a-1)}}
\left({2an_0 + \a + 1 \over 2} ~,~ {2an_0 + \b + 1 \over 2} ;
{am + \a + 1 \over 2} ~,~ {am + \b + 1 \over 2} \right)
}
Note that, as in the case of $a$ and $d$ prime, these states
can not have $\ell^+ \simeq N/2$ and $\ell^- \simeq 1$, which would
make them visible in the supergravity spectrum.
In order to describe the set of values $(s_1, s_2)$ more explicitly,
we need to analyze the congruence of \pppmn\ mod 2 and mod 4.


\newsec{Computation of the index for the $U(2)_r$ theories}

We will now consider a generalization to the theories
\eqn\youtooi{
Sym^N(\CS_r)
}
where $\CS_r$ is a product of a   bosonic WZW $U(2)_r$
with 4 MW fermions. As discovered in \refs{\Ivanov,\SevrinEW}
this theory has $\cag$ symmetry with
$(k^+,k^-) = (r+1,1) $
and $k=r+2$.
In this section we present some
partial results on the index for this theory.

The R-sector supercharacter before GSO projection is
\eqn\youtooii{
Z = \sum_{j=0}^{r/2} \biggl\vert 
\chi_j^{(r)}(\tau, \omega_+)\SCh^{\CS,R}(\tau,\omega_+,\omega_-) \biggr\vert^2
\sum_{\Gamma^{1,1}} q^{\half p_L^2} \bar q^{\half p_R^2}
}

It is worth computing the index first for the theory $\CS_r$. Using \schr,
identity (D.7) and the formula for Kac-Moody characters:
\eqn\kmchr{
\chi^{(r)}_j(\omega, \tau) = 
{ \Theta_{2j+1, r+2}^-(\omega, \tau)\over \Theta^-_{1,2}(\omega,\tau) }
}
we immediately find for the left-index:
\eqn\leftinx{
I_1 (\CS_r)
 = \sum_{j=0}^{r/2} \Theta_{2j+1, r+2}^-(\omega, \tau) 
\overline{\Xi_j} \sum_{\Gamma^{1,1}} q^{\half p_L^2}
\bar q^{\half p_R^2}
}
where in the charge zero sector
\eqn\youtooiii{
\Xi_\mu = \chi_j^{(r)}(\tau, \tilde \omega_+)
\SCh^{\CS,R}(\tau,\tilde \omega_+,\tilde \omega_-)
}
for $\mu=2j+1$.

Consider the term corresponding to $j$ in \leftinx.
Now, as usual, we have to worry that
  the index in principle receives contributions 
from any representation $(\l^+,\l^-)$
with
 $2 (\l^+ + \l^-) -1 = 2j+1$
on the left. Therefore, we need to 
examine the remaining character on the right.
Again applying the index to $\Xi_j$ we get $\Theta_{2j+1,r+2}^-$, so the
only massless representations which can occur in the expansion of
\youtooiii\ have $2 (\l^+ + \l^-) -1 = 2j+1$. 
Now we compute  $q$ expansions to get
\eqn\youtooiv{
\Xi_\mu = q^{(j+1/2)^2/(r+2)} (\tilde u - \tilde v) \chi_j(\tilde z_+) + \cdots
}
showing that only the representation $(j+1/2,1/2)$ actually appears.

It follows that the only massless representation 
on the right which appears is $(j+1/2,1/2)$.
Therefore, the simplest consistent solution is
\eqn\youtoov{
Z = \sum_{j=0}^{r/2} \biggl\vert \SCh(j+1/2,1/2) \biggr\vert^2 + \cdots
}
where the dots stand for massive representations.

Thus the RR BPS spectrum of the theory $\CS_r$ is
\eqn\antosp{
(j+1/2,1/2; j+1/2,1/2) \qquad  0 \leq j \leq r/2
}
Under spectral flow to   the NSNS sector we get
$(j,0;j,0) $  for $ 0 \leq j \leq r/2 $.

Let us now turn to the symmetric product \youtooi.
A small computation shows that   the left-index is
\eqn\leftindx{
I_1 (Sym^N(\CS_r)) = - z_+ {d\over dz_-}(T_N Z) =
\sum_{a\vert N} \sum_{n_0=0}^{d-1} \sum_{j=0}^{r/2} \Theta_{\mu,K}(\omega,\tau)
\overline{ \Xi_{j,n_0} }
}
where
\eqn\myukay{
\eqalign{
\mu & = a (2n_0(r+2) + 2j+1) \cr
K & = N (r+2) \cr}
}
\eqn\leftyout{
\Xi_{j,n_0} := {1\over d} \sum_{b=0}^{d-1} 
e^{-2\pi i {b\over d} (r+2)\bigl(n_0 + {2j+1\over 2r+4}\bigr)^2}
\biggl(\chi^{(r)}_j 
\SCh^{\CS,R}\biggr)({a\tau+b \over d}, a \tilde \omega_+, a \tilde \omega_-)
Z_{\Gamma}({a\tau+b \over d})
}

Taking another derivative to compute the index $I_2$ we find, in the
charge zero sector the projection operator constraining
constraint $\tl \mu = a ((2r+4) m_0 + 2j+1) = 2(\tl \l^+ + \tl \l^-) -1 $ is
\eqn\projct{
(m_0 - n_0 ) \biggl( (r+2) (m_0 + n_0) + 2j + 1\biggr) = 0 ~\mod~ d
}
This generalizes \projector.

Again, we can analyse the terms $a=1, d=N$ and $a=N,d=1$ as before.

\subsec{Analysis of $a=N, d=1$}

Now we attempt to decompose
\eqn\andone{
\chi^{(r)}_j \SCh^{\CS,R}(N\tau, N \omega_+, N\omega_-)
}
in terms of massless characters.
We note that the $q$-expansion of the $\CS_r$ theory  begins as
\eqn\adnonei{
\chi^{(r)}_j \SCh^{\CS,R}(\tau, \omega_+, \omega_-) =
e^{2\pi i \tau {(j+1/2)^2 \over r+2} } 
\chi_j(z_+) (\chi_{1/2}(z_+) - \chi_{1/2}(z_-) ) + \cdots
}

Now, let us work under the hypothesis that the only BPS states which
contribute come from this leading term in the $q$ expansion.
It then follows from the BPS bound that the only spins which
appear satisfy:
\eqn\oneadin{
\l^+ + \l^- = Nj + {N+1\over 2}
}

It is not difficult to  prove the generalization of \keysint:
\eqn\keysintp{
\eqalign{
&
\chi_j(z_+^N)\bigl(
\chi_{1/2}(z_+^N) - \chi_{1/2}(z_-^N)\bigr)  = (u-v) \sum_{a + b = N-1}
\biggl[ \chi_{Nj + a/2}(z_+) \chi_{b/2}(z_-) \cr
&\quad\quad - \chi_{Nj + (a-1)/2}(z_+) \chi_{(b-1)/2}(z_-) \biggr]
+ (u-v)^2 \chi_{R}(z_+,z_-) \cr}
}
where the second term has a second order zero at $u=v$ and
corresponds to the leading  $q$-expansion of a massive character.
Therefore, we conclude that
\eqn\andonepp{
\chi^{(r)}_j \SCh^{\CS,R}(N\tau, N \omega_+, N\omega_-) = \sum_{a+b=N-1}
(-1)^{N-a} \SCh(  Nj + (a+1)/2,   (N-a)/2 ) + \cdots
}
where we are using supercharacters of $\cag$ at level
$(k^+ = N(r+1),k^- = N)$.
The simplest spectrum consistent with the index is
\eqn\rmadn{
\bigoplus_{j=0}^{r/2} \bigoplus_{a=0}^{N-1} 
\Bigl\vert (  Nj + (a+1)/2,  (N-a)/2 )\Bigr\vert^2.
}
Upon spectral flow to the NS sector this is
\eqn\rmadn{
\bigoplus_{j_1=0}^{r/2} \bigoplus_{j_2=0}^{\half( N-1)}
\Bigl\vert \bigl(Nj_1 + j_2, j_2\bigr)_{NS}\Bigr\vert^2.
}
Once again, we remind the reader that this spectrum is not unique;
one can add virtual representations of the net index zero.

\subsec{Analysis of $a=1, d=N$}

Now \leftyout\ simplifies to
\eqn\leftyout{
\Xi^{ls}_{j,n_0} = {1\over N} \sum_{b=0}^{N-1}
e^{-2\pi i {b\over N} (r+2)\bigl(n_0 + {2j+1\over 2r+4}\bigr)^2}
\biggl(\chi^{(r)}_j 
\SCh^{\CS,R}\biggr)({\tau+b \over N},  \tilde \omega_+,  \tilde \omega_-)
}
%
In order to expand this expression in terms of supercharacters,
first we would need to write it as a power series in $q$.
It is convenient to introduce
\eqn\srjdef{
S_j := \chi^{(r)}_j \SCh^{\CS,R}
= q^{{ (2j+1)^2 \over 4 (r+2)}} \sum_{n=0}^{\infty} q^n s_{j,n} (z_+ , z_-)
}
where $s_{j,n} (z_+ , z_-)$ are certain combinations
of $SU(2) \times SU(2)$ characters.
Then, as in the previous section, one can easily perform the sum
over $b$ in \leftyout. This leads to
\eqn\leftyoutb{
\Xi^{ls}_{j,n_0} = \sum_{{n=0 \atop
n = n_0 ( (r+2)n_0 + 2j + 1) ~\mod~N}}^{\infty}
q^{{1 \over N} \bigl( n + {(2j+1)^2 \over 4 (r+2)} \bigr)}
s_{j,n} (z_+ , z_-)
}
which is a generalization of \ximubh.
Unfortunately, the explicit calculation of $s_{j,n} (z_+ , z_-)$ is
very difficult for generic values of $r$ and $j = 0, \ldots, r/2$.
However, we analyzed \leftyoutb\ in a number of examples.
Curiously, in all the cases that we studied, $\Xi^{ls}_{j,n_0}$
can be expanded in terms of supercharacters with non-negative multiplicities.

\example{$N=5$, $r=2$, $j=1$, $n_0=2$}
\eqn\nlsexmplb{\eqalign{
\Xi^{ls}_{1,2} & = \SCh(9/2,3/2) + \cr
& + 2 \SChm (41/80, 5/2, 2/2) + \SChm (41/80, 4/2, 3/2) + \cr
& + 3 \SChm (41/80, 3/2, 2/2) + 3 \SChm (121/80, 9/2, 2/2) + \cr
& + 11 \SChm (121/80, 8/2, 3/2) + 7 \SChm (121/80, 7/2, 4/2) + \cr
& + \SChm (121/80, 6/2, 5/2) + 36 \SChm (121/80, 7/2, 2/2) + \cr
& + 63 \SChm (121/80, 6/2, 3/2) + 32 \SChm (121/80, 5/2, 4/2) + \cr
& + 3 \SChm (121/80, 4/2, 5/2) + 115 \SChm (121/80, 5/2, 2/2) + \cr
& + 123 \SChm (121/80, 4/2, 3/2) + 36 \SChm (121/80, 3/2, 4/2) + \cr
& + 2 \SChm (121/80, 2/2, 5/2) + 124 \SChm (121/80, 3/2, 2/2) + \cr
& + 66 \SChm (121/80, 2/2, 3/2) + \ldots
}}
\endexample


\bigskip
\noindent{\bf Acknowledgements:}
G.M. would like to that the KITP for hospitality during
the course of some of this work.
This work was conducted during the period S.G. served as a Clay
Mathematics Institute Long-Term Prize Fellow. The work of E.M. is
supported in part by DOE grant DE-FG02-90ER40560, that of G.M. by
DOE grant DE-FG02-96ER40949 and that of A.S. and S.G. by DE-FG02-91ER40654.


\appendix{\A}{Leading $q$ expansion of the Peterson-Taormina characters}

\subsec{ Massive R-sector characters}

The unitarity range in the R sector is\foot{\PTone\ eq. 2.9
and \PTtwo\ eq. 2.18 disagree on the upper limit. We  take \PTtwo\ eq. 2.18. }
\eqn\unitary{  1 \leq \l^\pm < \half k^\pm }
and massive representations satisfy:
\eqn\untry{
\eqalign{
h -c/24 & \geq u^2/k + (\l^+ + \l^--1)^2/k\cr}
}
%

In the unitarity range, the $q$ expansion is always:
\eqn\massvrp{
\SCh_{\rm massive}( h , \l^{\pm} )= (-1)^{2\l^-} q^{h-c/24} (\chi_{\half}(z_+ )
-  \chi_{\half}(z_-))^2 \chi_{\l^+-1}(z_+) \chi_{\l^--1}(z_-) +\ldots
}

\subsec{Massless R-sector characters}

The unitary range is now:
\eqn\unitaryrg{
%
%
\half  \leq \l^\pm \leq \half k^\pm
}
%
%
\eqn\untryp{
\eqalign{
h -c/24 & = u^2/k + (\l^+ + \l^-- 1/2 )^2/k\cr}
}
%

For the massless characters we find the leading term is given by
%
\eqn\msslessi{
\eqalign{
&
\SCh(\l^+,\l^-)= (-1)^{2\l^--1}
q^{h-c/24} \cdot \cr
&\cdot (\chi_{\half}(z_+ ) -  \chi_{\half}(z_-))
\biggl( \chi_{\l^+-1/2}(z_+) \chi_{\l^--1/2}(z_-) -
\chi_{\l^+- 1 }(z_+) \chi_{\l^- - 1 }(z_-) \biggr) + \ldots \cr}
}
Here it is understood that   $\chi_{-1/2}=0$.

\subsec{Massive NS-sector characters}

Unitarity bounds are:\foot{Again, the lower limit
differs from \PTone, eq. 2.10.  Their statement
is not consistent with spectral flow.}
\eqn\untip{
0 < \l^\pm <  \half(k^\pm -1)
}
with
\eqn\boudn{
kh > (\l^+-\l^-)^2 + k^- \l^+ + k^+ \l^- + u^2
}
Note this region is empty for $k^\pm =1$. In that case there are only
massless characters.

%

The leading term in the $q$ expansion is:
\eqn\leading{
\SCh^{\CA_\gamma, NS}_{\rm massive}
= (-1)^{2\l^-} q^{h - c/24} \chi_{\l^+}(z_+) \chi_{\l^-}(z_-) + \ldots
}

\subsec{Massless NS-sector characters}

Unitarity bounds are:
\eqn\untip{
0 \leq \l^\pm \leq \half(k^\pm -1)
}
with
\eqn\boudn{
kh = (\l^+-\l^-)^2 + k^- \l^+ + k^+ \l^- + u^2
}
The leading term in the $q$ expansion
is the {\it same} for the massless and massive case and is:

\eqn\msslsns{
\SCh^{\CA_\gamma, NS}(\l^+,\l^-) 
= (-1)^{2\l^-} q^{h-c/24} \chi_{\l^+}(z_+ ) \chi_{\l^-}(z_-) + \cdots
}

The massless and massive characters differ at $q^{h+1/2}$.


\appendix{B}{Characters of the Large $\CN=4$ SCA}

The character of a {\it massive} representation in
the Ramond sector of the $\CA_{\g}$ is given by \PTone:
\eqn\agchar{\eqalign{
{\rm Ch}^{\CA_{\g}, R} & (k^{\pm} , h , \l^{\pm} ; q, z_{\pm})
 = q^{h-c/24} \left( F^R (q,z_+, z_-) \right)^2
(1+ z_+^{-1} z_-^{-1})^2 (1+ z_+^{-1} z_-)^2 \cr
& \times \prod_{n=1}^{\infty}
(1-q^n)^{-4} (1 - z_+^2 q^n)^{-1} (1 - z_+^{-2} q^{n-1})^{-1}
(1 - z_-^2 q^n)^{-1} (1 - z_-^{-2} q^{n-1})^{-1}  \cr
& \times \sum_{m=-\infty}^{+\infty} q^{m^2k^+ + (2\l^+ - 1)m}
\Big( z_+^{2(mk^+ + \l^+)} - z_+^{-2(mk^+ + \l^+ -1)} \Big) \cr
& \times \sum_{n=-\infty}^{+\infty} q^{n^2 k^- + (2\l^- - 1)n}
\Big( z_-^{2(nk^- + \l^-)} - z_-^{-2(nk^- + \l^- -1)} \Big)
}}
where
\eqn\frdef{F^R (q,z_{\pm})
= \prod_{n \in \Z_{+}}
(1 + z_+ z_- q^n)(1 + z_+^{-1} z_-^{-1} q^n)
(1 + z_+ z_-^{-1} q^n)(1 + z_+^{-1} z_- q^n)
}

The authors of \PTone\PTtwo\ mostly work with $\tilde \CA_{\g}$
which is obtained from $\CA_{\g}$ by removing a free boson and
four fermions. The quantum numbers of representations of these
two algebras are related via, see {\it e.g.} \GPTVP,
\eqn\avsatilde{
\matrix{
\underline{\CA_{\g}} & & \underline{\tilde \CA_{\g}} \cr
c & \to & \tilde c = c-3 \cr
k_{\pm} & \to & ~~~~\tilde k_{\pm} = k_{\pm} - 1 \cr
\l^{\pm}_{NS} & \to & ~~\tilde \l^{\pm}_{NS} = \l^{\pm}_{NS} \cr
\l^{\pm}_R & \to & ~~~~\tilde \l^{\pm}_R = \l^{\pm}_R - {1 \over 2}
}}
and, similarly, for the characters:
\eqn\chaa{
{\rm Ch}^{\CA_{\g}, R} (k^{\pm} , h , \l^{\pm} ; q, z_{\pm})
= {\rm Ch}^{\CS, R} (q, z_{\pm})
\times {\rm Ch}^{\tilde \CA_{\g}, R}
(\tilde k^{\pm} , \tilde h , \tilde \l^{\pm} ; q, z_{\pm})
}
where
\eqn\schar{{\rm Ch}^{\CS, R} (q, z_{\pm})
= q^{u^2/k + 1/8} F^R (q, z_{\pm})
\times \prod_{n=1}^{\infty} (1 - q^n)^{-1}
(1 + z_+^{-1} z_-^{-1})(1 + z_+^{-1} z_-) z_+
}

Notice, that \schar\ vanishes when $z_+ = - z_-$.
{}From \chaa\ it follows that ${\rm Ch}^{\CA_{\g}, R}$
also vanishes, as long as ${\rm Ch}^{\tilde \CA_{\g}, R}$
is finite at $z_+ = - z_-$.

In   \PTtwo, Petersen and Taormina
give
an independent derivation of the $\tilde \CA_{\g}$ characters for
massive states,
\eqn\tilachar{\eqalign{
{\rm Ch}^{\tilde \CA_{\g}, R} & (\tilde k^{\pm} , h , \l^{\pm} ; q, z_{\pm})
= q^{h-\tilde c/24} F^R (q,z_+, z_-) \cr
& \times \prod_{n=1}^{\infty}
(1-q^n)^{-2} (1-z_+^2 q^n)^{-1} (1-z_+^{-2} q^n)^{-1}
(1-z_-^2 q^n)^{-1} (1-z_-^{-2} q^n)^{-1} \cr
& \times \prod_{n=1}^{\infty} (1-q^n)^{-1}
(z_+^{-1} + z_-^{-1}) (1 + z_+^{-1} z_-^{-1})
(1 - z_+^{-2})^{-1} (1 - z_-^{-2})^{-1} \cr
& \times \sum_{m,n=-\infty}^{\infty} q^{n^2k^+ + 2 \l^+ n + m^2 k^- + 2\l^- m}
\sum_{\epsilon_+,\epsilon_- = \pm 1}
\epsilon_+ \epsilon_-
z_+^{2\epsilon_+ (\l^+ + nk^+)}
z_-^{2\epsilon_- (\l^- + mk^-)}
}}
which they claim agrees with \chaa.
For example, it is easy to check that both \tilachar\ and \schar\
have a simple zero at $z_+ = - z_-$, whereas \agchar\ has a double zero.

For {\it massless} states the characters look like \PTtwo:
\eqn\zerochar{\eqalign{
{\rm Ch}_0^{\tilde \CA_{\g}, R} & (\tilde k^{\pm} , h , \l^{\pm} ; q, z_{\pm})
= q^{h-\tilde c/24} F^R (q,z_+, z_-) \cr
& \times \prod_{n=1}^{\infty}
(1-q^n)^{-2} (1-z_+^2 q^n)^{-1} (1-z_+^{-2} q^n)^{-1}
(1-z_-^2 q^n)^{-1} (1-z_-^{-2} q^n)^{-1} \cr
& \times \prod_{n=1}^{\infty} (1-q^n)^{-1}
(z_+^{-1} + z_-^{-1}) (1 + z_+^{-1} z_-^{-1})
(1 - z_+^{-2})^{-1} (1 - z_-^{-2})^{-1} \cr
& \times \sum_{m,n=-\infty}^{\infty} q^{n^2k^+
+ 2 \l^+ n + m^2 k^- + 2\l^- m} \cr
& \times \sum_{\epsilon_+,\epsilon_- = \pm 1}
\epsilon_+ \epsilon_-
z_+^{2\epsilon_+ (\l^+ + nk^+)}
z_-^{2\epsilon_- (\l^- + mk^-)}
(z_+^{-\epsilon_+} q^{-n} + z_-^{-\epsilon_-} q^{-m})^{-1}
}}
Using \chaa\ we can obtain the character of the corresponding
massless state in the Ramond sector in the $\CA_{\g}$ theory.
Since \zerochar\ is finite at $z_+ = - z_-$,
the $\CA_{\g}$ character of a massless state
has only a simple zero at this point.

For completeness, we include here the  characters
${\rm Ch}^{\CN=4} (k,\ell;q,z)$ of the small $\CN=4$ superconformal
algebra:
\eqn\smallchar{\eqalign{
{\rm Ch}^{\CN=4,R} (k,\ell)
& = z q^{k/4}
\prod_{n=1}^{\infty}
{ (1+z q^n)^2 (1+z^{-1} q^{n-1})^2 \over
(1-q^n)^{2} (1-q^n z^2) (1-q^{n-1} z^{-2}) } \cr
& \times \sum_{m \in \Z}
q^{(k+1)m^2 + 2 \ell m}
\Big[ {z^{2m(k+1)+2 \ell -1} \over (1+z^{-1} q^{-m})^2 }
- {z^{-2m(k+1)-2 \ell-1} \over (1+z q^{-m})^2 } \Big]
}}
%


\appendix{C}{Spectral flow rules}

The spectral flow isomorphism involves, among other things:
\eqn\sfi{ \eqalign{ & \hat L_0 = L_0 - \rho A_{0}^{+,3} + {1\over
4} k^+ \rho^2 \cr & \hat A_{0}^{+,3} = A_{0}^{+,3} - \half \rho
k^+  \cr & \hat A_{0}^{-,3} = A_{0}^{-,3}   \cr} }
We will call this $s_+^\rho$. Notice $\rho$ has to be an integer
since $A_{0}^{+,3}$ should have half-integer spectrum. If $\rho$
is odd it exchanges NS and R sectors.

There is the corresponding flow on the other $SU(2)$ called $s_-^\eta$.
Among other things:
\eqn\sfi{ \eqalign{ & \hat L_0 = L_0 - \eta A_{0}^{-,3} + {1\over
4} k^- \eta^2 \cr & \hat A_{0}^{+,3} = A_{0}^{+,3}  \cr & \hat
A_{0}^{-,3} = A_{0}^{-,3}- \half \eta k^-   \cr} }
Spectral flow by $\rho=1$ or $\eta=1$ defines operators
$s_\pm $ exchanging $R$ and $NS$ sector representations.
The behavior of the highest weight states under spectral
flow is complicated, and hence it
is important to distinguish the action of $s_\pm$ on states
and on representation labels. On the latter we have:

\bigskip
\noindent
On massive representations:
\eqn\massvflw{
s_+: (h,\l^+,\l^-)_{NS} 
\leftrightarrow (h-\l^+ + {1\over 4} k^+, \half k^+ - \l^+, \l^-+1)_R
}
On massless representations:
\eqn\massless{
\eqalign{
 s_+:
& (h,\l^+,\l^-)_{NS} \rightarrow 
(h-\l^+ + {1\over 4} k^+, \half k^+ - \l^+, \l^-+ \half )_R\cr
& (h,\l^+,\l^-)_R \rightarrow 
(h-\l^+ + {k^+\over 4} , \half k^+ - \l^+, \l^- - \half )_{NS} \cr}
}
\eqn\massless{
\eqalign{
 s_-:
& (h,\l^+,\l^-)_{NS} \rightarrow 
(h-\l^- + {1\over 4} k^-,   \l^+ +\half , \half k^- - \l^-  )_R\cr
& (h,\l^+,\l^-)_R \rightarrow 
(h-\l^- + {k^-\over 4} ,   \l^+ -\half , \half k^- - \l^-   )_{NS} \cr}
}
%


\appendix{D}{Holomorphic level $k$ theta functions}

Our convention for the level $1/2$ theta functions is:
\eqn\infprdct{
\eqalign{
\vt{\theta}{\phi}{0}
& := \sum q^{\half (n+ \theta)^2 } e^{ 2 \pi i (n+\theta)\phi}  \cr
 = e^{ 2 \pi i \theta \phi}
q^{{\theta^2 \over  2}  }
&
\prod_{n=1}^\infty (1-q^n)(1+ e^{2 \pi i \phi} q^{n-\half + \theta} )
(1+ e^{- 2 \pi i \phi} q^{n-\half -\theta} ) \cr
}
}

The standard definition of the special theta functions is:
\eqn\thtprd{
\eqalign{
\vartheta_1(\omega\vert \tau)& := \vt{1/2}{1/2}{\omega}   \cr
}
}

It has a product representation:
\eqn\spcthtprd{
\eqalign{
\vartheta_1(\omega\vert \tau)& = -2 \sin(\pi \omega) q^{1/8} \prod_{n=1}^\infty
(1-q^n)(1- e^{2\pi i \omega}q^n )(1- e^{-2\pi i \omega}q^n )\cr
}
}

Level $k$ theta functions
$\Theta_{\mu,k}(\omega,\tau)$, $\mu=-k+1, \dots, k$
 are defined by:
\eqn\thetmk{
\eqalign{
\Theta_{\mu,k}(\omega,\tau)
& = \sum_{\ell\in \IZ, \ell = \mu ~\mod~ 2k } q^{\ell^2/(4k)} y^{\ell} \cr
& \equiv \sum_{n\in \IZ} q^{k(n+\mu/(2k))^2}
y^{ (\mu + 2k n)} \cr
& = q^{{\mu^2 \over  4 k}} y^{\mu}
\sum_{n\in \IZ} q^{k n^2 + n\mu }
y^{2 kn } \cr}
}
where $y=e^{2\pi i \omega}$.

There are $2k$ independent functions which may be split into $(k+1)$ even and
$(k-1)$ odd functions of $\omega$.

Note:
\eqn\simpids{
\eqalign{
\Theta_{\mu, k }(\omega,\tau) & = \Theta_{\mu + 2k a,k}(\omega,\tau)
 \qquad a\in \IZ \cr
\Theta_{\mu, k }(-\omega,\tau) & = 
\Theta_{2k-\mu ,k}(\omega,\tau)= \Theta_{-\mu,k}(\omega,\tau) \cr}
}

It is useful to define:
\eqn\dfthem{
\Theta^\pm_{\mu,k}(\omega, \tau):= 
\Theta_{\mu}(\omega,\tau) \pm  \Theta_{\mu}(-\omega,\tau)
}

Then:
\eqn\relthetone{
\Theta_{1,2}(\omega,\tau) - \Theta_{-1,2}(\omega,\tau) =
-i \vartheta_1(2 \omega \vert \tau)
}

To compute Hecke transforms we need
\eqn\heckei{
 \Theta_{\mu,k}\bigl(a \omega, {a \tau + b\over d}\bigr)
= \sum_{n_0=0}^{d-1} e^{2\pi i {b\over d}k (n_0 + \mu/(2k))^2}
\Theta_{a(2kn_0 + \mu), Mk}(\omega,\tau)
}
where $M=ad$. (There is no need to assume $a,d$ are relatively prime.)
Notice that one can symmetrize or anti-symmetrize \heckei\ under
$\omega \to - \omega$.

Modular transformations:
\eqn\tauplone{
\Theta_{\mu,k}(\omega, \tau+1) = 
e^{2\pi i {\mu^2\over 4k}} \Theta_{\mu,k}(\omega,\tau)
}
\eqn\tauotau{
\Theta_{\mu,k}(-\omega/\tau ,-1/ \tau)= 
(-i \tau)^{1/2} e^{2\pi i k \omega^2/\tau}
\sum_{\nu=0}^{2k-1} {1\over \sqrt{2k}} 
e^{2\pi i {\mu\nu\over 2k} } \Theta_{\nu,k}(\omega,\tau)
}
It is useful to note the projection onto odd theta functions:
\eqn\tauotau{
\Theta_{\mu,k}^-(-\omega/\tau ,-1/ \tau)= 
i (-i \tau)^{1/2} e^{2\pi i k \omega^2/\tau}
\sum_{\nu=1}^{k-1} \sqrt{2\over k} 
\sin (\pi {\mu\nu\over k} ) \Theta_{\nu,k}^-(\omega,\tau)
}
%


\appendix{E}{Siegel-Narain  Theta functions}

Let $\Lambda$ be a lattice of signature $(b_+,b_-)$.  Let $P$
be a decomposition
of $\Lambda\otimes \IR$ as a sum of orthogonal subspaces
 of definite
signature:
\eqn\dfsign{ P:\Lambda \otimes \IR \cong
\IR^{b_+,0} \perp \IR^{0,b_-}
}
Let $P_\pm(\lambda)= \lambda_\pm $ denote the projections onto the two
factors.
We also write $\lambda = \lambda_+ + \lambda_-$.
 With our conventions $P_-(\lambda)^2 \leq 0 $.

Let  $\Lambda+ \gamma $ denote a translate of the lattice
$\Lambda$.
We define the Siegel-Narain theta function
\eqn\sglthet{
\eqalign{
\Theta_{\Lambda + \gamma} (\tau, \alpha,\beta; P, \xi)
\equiv
&
 \exp[{ \pi \over  2 y} ( \xi_+^2 - \xi_-^2) ] \cr
\sum_{\lambda\in \Lambda + \gamma}
\exp\biggl\{ i \pi \tau (\lambda+ \beta)_+^2 +
i \pi \bar \tau (\lambda+ \beta)_-^2
&
+ 2 \pi i (\lambda+\beta, \xi) - 2 \pi i
(\lambda+\half \beta, \alpha) \biggr\} \cr
= & e^{i \pi (\beta,\alpha)}
 \exp[{ \pi \over  2 y} ( \xi_+^2 - \xi_-^2) ] \cr
\sum_{\lambda\in \Lambda + \gamma}
\exp\biggl\{ i \pi \tau (\lambda+ \beta)_+^2 +
i \pi \bar \tau (\lambda+ \beta)_-^2
&
+ 2 \pi i (\lambda+\beta, \xi) - 2 \pi i
(\lambda+  \beta, \alpha) \biggr\} \cr}
}
where $y= \tau_2$.

The main transformation law is:
\eqn\thetess{
\Theta_{\Lambda } (-1/\tau, \alpha,\beta; P, {\xi_+ \over  \tau} +
{\xi_- \over  \bar \tau} )
= \sqrt{\vert \Lambda \vert \over  \vert \Lambda^* \vert}
(-i \tau)^{b_+/2} (i \bar \tau)^{b_-/2}
\Theta_{\Lambda' } ( \tau, \beta,-\alpha ; P, \xi )
}
where $\Lambda'$ is the dual lattice.
If there is a characteristic vector, call it $w_2$, such that
\eqn\characteristic{
(\lambda,\lambda) = (\lambda, w_2)~ \mod ~2
}
for all $\lambda$
then we have in addition:
\eqn\thettee{
\Theta_{\Lambda } (\tau+1, \alpha,\beta; P, \xi)
= e^{-i \pi(\beta,w_2)/2}
\Theta_{\Lambda } ( \tau, \alpha-\beta-\half w_2,\beta ; P, \xi )
}


\appendix{F}{Symmetric product partition functions}

We review the derivation of \DijkgraafXW.
The Hilbert space is
\eqn\symmprdrr{
\CH({\rm Sym}^N(\CC_0)) = \bigoplus_{(n)^{\l_n} } \bigotimes_n
{\rm Sym}^{\l_n}(\CH^{(n)}(\CC_0))
}
where we sum over partitions $\sum n \l_n = N$.

So the generating function is
\eqn\genfn{
\CZ := 1 + \sum_{N\geq 1} 
p^N {\Tr}_{\CH({\rm Sym}^N(\CC_0))} q^H y^J \bar q^{\bar H} \bar y^{\bar J}
= \prod_{n=1}^\infty 
\sum_{\ell_n=0}^\infty p^{n \ell_n} {\Tr}_{{\rm Sym}^{\l_n}(\CH^{(n)} )}
q^H y^J \bar q^{\bar H} \bar y^{\bar J}
}
Here $y^J$ is short for insertions of $z_+, z_-$ etc.

Now the standard formula for traces in 
symmetric products of vector spaces gives
\eqn\symmtrp{
\sum_{\ell_n=0}^\infty p^{n \ell_n} {\Tr}_{{\rm Sym}^{\l_n}(\CH^{(n)} )}
q^H y^J \bar q^{\bar H} \bar y^{\bar J}
= \prod_{{\rm basis}~\CH^{(n)}} 
{1\over 1- p^n q^H y^J \bar q^{\bar H} \bar y^{\bar J} }
}
where we take a product over an eigenbasis in 
$\CH^{(n)}(\CC_0)$, the Hilbert space of
a string of length $n$.  But
\eqn\longstr{
{\Tr}_{\CH^{(n)}(\CC_0) }( q^H y^J \bar q^{\bar H} \bar y^{\bar J} )
= {1\over n} \sum_{b=0}^{n-1} 
{\Tr}_{\CH(\CC_0) } \omega^b q^{{1\over n}H} y^J \bar q^{{1\over n}\bar H}
\bar y^{\bar J}
}
where $\omega= e^{2\pi i (L_0-\bar L_0)/n}$, and $H= L_0 - c/24$.
Thus the sum on $b$ projects to states that 
satisfy $\Delta - \bar \Delta = 0 ~\mod~ n$.
{}From this the symmetric product formula follows.

If we have a $\IZ_2$-graded Hilbert space then we 
should take a supertrace, and
use the rule:
\eqn\gensumst{
\sum_{\ell =0}^\infty p^\ell \STr_{{\rm Sym}^\ell(\CH)}(\CO) =
\prod_{{\rm eigenbasis}~\CH_0} {1\over (1-p \CO_i) }
\prod_{{\rm eigenbasis}~\CH_1} (1-p\CO_i) =
\exp\biggl[ \sum_s {p^s \over s} \STr_{\CH}(\CO^s)\biggr]
}%

On the other hand, for the Hecke operator formula we take the logarithm
of \genfn\ using \symmtrp:
\eqn\logfrm{
\log \CZ = \sum_{n=1}^\infty \sum_{s=1}^\infty
\sum_{{\rm basis}~\CH^{(n)}} {1\over s} p^{ns} 
(q^H y^J \bar q^{\bar H} \bar y^{\bar J})^s
}
Using again \longstr\ this can be written as
\eqn\heckeform{
\log \CZ = \sum_{N=1}^\infty p^N T_N Z_0
}
where
\eqn\teenn{
T_N Z_0 := {1\over N}  
\sum_{s=1, s\vert N}^N \sum_{b=0}^{n-1} 
Z_0({s \tau + b \over n}, y^s; {s \bar \tau + b \over n} ,
\bar y^s)
}
and $n=N/s$.
Using \gensumst\ we see that this also holds for the case of the supertrace.


\appendix{G}{Examples of the index for iterated symmetric products}

The partition function of the symmetric product theory Sym$^N (\CS)$,
with $N = Q_1 Q_5$, can be written in the compact form \polyexp:
\eqn\zsym{
Z \Big( {\rm ~Sym}^N (\CS) \Big) = T_N Z_0 + \ldots
}
where $T_N$ denotes the Hecke operator, and the dots stand for the
higher-order terms which do not contribute to the index (see
section 4). Similarly, the partition function of the iterated
symmetric product theory looks like
\eqn\zsymsym{
Z \Big( {\rm ~Sym}^{Q_1} {\rm ~Sym}^{Q_5} (\CS) \Big)
= T_{Q_1} T_{Q_5} Z_0 + \ldots
}
Our goal here is to compare the massless spectrum
of these two theories.
Using the properties of the Hecke operators summarized in section
4.2, we can write the linear term in \zsym\ as
\eqn\zsymp{
T_N Z_0 = \prod_p T_{p^{e_p}} Z_0
}
where $N = Q_1 Q_5 = \prod p^{e_p}$.
Similarly, factorizing $Q_1 = \prod p^{e_p'}$ and $Q_5 = \prod p^{e_p''}$,
we have
\eqn\zsymsymp{
T_{Q_1} T_{Q_5} Z_0 = \prod_p T_{p^{e_p'}} T_{p^{e_p''}} Z_0
}
In particular, from \ahecksi\ it follows that, when $Q_1$ and $Q_5$
are relatively prime, the partition function \zsymp\ of the symmetric
product theory is equal to the partition function \zsymsymp\
of the iterated symmetric product theory.
Hence, in order to see a difference between these two theories,
one should consider $Q_1$ and $Q_5$ which are {\it not} relatively prime.

In order to see the difference between \zsymp\ and \zsymsymp\
when the prime factor $p$ occurs in both $Q_1$ and $Q_5$,
it is instructive to consider a simple case
\eqn\qqpr{Q_1 = p^{r_1} \quad , \quad Q_5 = p^{r_2}}
so that
\eqn\nprr{N = p^r \quad , \quad r = r_1 + r_2}
In this case, in order to evaluate the difference\foot{Again,
the dots stand for the higher order terms that do not contribute
to the index.}
%
\eqn\zznpr{
Z \Big( {\rm ~Sym}^{Q_1} {\rm ~Sym}^{Q_5} (\CS) \Big)
- Z \Big( {\rm ~Sym}^{Q_1 Q_5} (\CS) \Big)
= T_{p^{r_1}} T_{p^{r_2}} Z_0 - T_{p^r} Z_0 + \ldots}
we need to find a relation between the Hecke operators
$T_{p^{r_1}} T_{p^{r_2}}$ and $T_{p^r}$.
Using \primpow, we find the desired identity
\eqn\ttviat{
T_{p^{r_1}} T_{p^{r_2}} = \sum_{k=0}^{\min (r_1 , r_2)}
{1 \over p^k} T_{p^{r_1 + r_2 - 2k}} W_{p^k}
}
where we also used multiplicativity of $W_p$, $W_p^k = W_{p^k}$.

%

Now, substituting \ttviat\ into \zznpr, we obtain
\eqn\zznprt{\eqalign{
&
T_{p^{r_1}} T_{p^{r_2}} Z_0 (\tau , z_{\pm} )
- T_{p^r} Z_0 (\tau , z_{\pm} )
= \cr
& ~~~~~~~~~= \sum_{k=1}^{\min (r_1 , r_2)}
{1 \over p^k} T_{p^{r - 2k}} W_{p^k} Z_0 (\tau , z_{\pm} )
= \sum_{k=1}^{\min (r_1 , r_2)}
{1 \over p^k} T_{p^{r - 2k}} Z_0 (\tau , z_{\pm}^{p^k} ) = \cr
& ~~~~~~~~~= \sum_{k=1}^{\min (r_1 , r_2)}
{1 \over p^k} {1 \over p^{r-2k}} \sum_{ad = p^{r-2k}}
\sum_{b=0}^{d-1}
Z_0 \left( {a \tau + b \over d} , z_{\pm}^{ap^k} \right) = \cr
& = \sum_{k=1}^{\min (r_1 , r_2)}
{1 \over p^{r-k}} \sum_{\d = 0}^{r-2k} \sum_{b=0}^{p^{\d}-1}
Z_0 \left( p^{r-2k-2\d} \tau + {b \over p^{\d}} , z_{\pm}^{p^{r-k-\d}} \right)
}}

Therefore, the problem reduces to evaluating the terms of the form
$T_{p^{r - 2k}} Z_0 (\tau , z_{\pm}^{p^k} )$, which is similar to
the problem studied in section 5.8. In particular, among the
massless states of the iterated symmetric product theory ${\rm
~Sym}^{Q_1} {\rm ~Sym}^{Q_5} (\CS)$, which are not contained in
the ${\rm ~Sym}^{Q_1 Q_5} (\CS)$ theory, we find the states of the
form \specp\ where $a=p^{r-k-\d}$:
\eqn\ttstates{
\bigoplus_{{(n_0, k, \d, s) \atop \a + \b = (a-1)}}
\left({2an_0 + \a + 1 \over 2} ~,~ {2an_0 + \b + 1 \over 2} ;
{am + \a + 1 \over 2} ~,~ {am + \b + 1 \over 2} \right)
}
As in \mnsol\ -- \nrun, there are two families of solutions
for $n_0$ and $m$:
\eqn\ttmnone{\eqalign{
& 1) \quad \quad  m=2n_0 + sp^{\d} 
\quad\quad s=-1,0,1, \ldots, p^{r-2k-\d}-1  \cr
& ~~ \quad \quad \max \left( 0 , -{sp^{\d} \over 2}\right)
\le n_0 \le \min \left( p^{\d}-1 , {(p^{r-2k-\d}-s)p^{\d} -1 \over 2}\right)
}}
and
\eqn\ttmntwo{\eqalign{
& 2) \quad \quad  m=sp^{\d}-2n_0 -1 \quad\quad s=1,2, \ldots, p^{r-2k-\d}+1 \cr
& ~~ \quad \quad \max \left( 0 , {sp^{\d} - p^{r-2k} \over 2}\right)
\le n_0 \le \min \left( p^{\d}-1 , {sp^{\d} -1 \over 2}\right)
}}
In both cases, $k$ runs from 1 to $\min (r_1 , r_2)$
and $\d = 0, \ldots, (r-2k)$, {\it cf.} \zznprt.
Note, that none of these extra states, which appear
when ${\rm g.c.d.} (Q_1,Q_5) >1$, can be a supergravity state.

\example{$Q_1 = Q_5 = p$}
Consider a simple example, where $r_1 = r_2 = 1$, so that $r=2$.
In this example, we have $k=1$ and $\d=0$, so that only the term
$Z_0 (\tau , z_{\pm}^{p^2} )$ appears on the right-hand side of \zznprt.
Moreover, the two families of solutions, \ttmnone\ and \ttmntwo,
collapse to single solution
\eqn\ppmn{ m=n_0=0 }
Therefore, the spectrum \ttstates\ takes the following simple form:
\eqn\ttstatespp{
\bigoplus_{\a + \b = p-1}
\left({\a + 1 \over 2} , {\b + 1 \over 2} ;
{\a + 1 \over 2} , {\b + 1 \over 2} \right)_{R}
}
After the spectral flow to the NS sector, we get
\eqn\ttstatesppns{
\bigoplus_{\b = 0}^{p-1}
\left( {p(p-1) - \b \over 2} ~,~ {\b \over 2} ~;~
{p(p-1) - \b \over 2} ~,~ {\b \over 2} \right)_{NS}
}
\endexample

In addition, in general there are extra states \extrastates\ coming from
non-trivial factorization of \mnnpr. The values of $n_0$ and $m$
for these states are given by \pppmn, where $s_1$ and $s_2$
assume their values in the range \ssbound. As we discussed above,
there are no additional states for $p=2$.
In this case, \ttstates\ -- \ttmntwo\ is the complete answer.

\listrefs

\bye